\documentclass[prb,twocolumn,amsmath,amssymb,showpacs]{revtex4}
\usepackage{color}
\usepackage{graphicx}
\usepackage{epsfig}
\usepackage{dcolumn}
\usepackage{bm}

\begin{document}
\title{Josephson photonics with a two-mode superconducting circuit}
\author{A. D. Armour$^1$, B. Kubala$^2$, and J. Ankerhold$^2$}
\affiliation{$^1$ School of Physics and Astronomy, University of Nottingham, Nottingham NG7 2RD, UK\\
$^2$ Institute for Complex Quantum Systems, University of Ulm, 89069 Ulm, Germany}


\begin{abstract}
We analyze the quantum dynamics of two electromagnetic oscillators coupled in series to a voltage biased Josephson junction. When the applied voltage leads to a Josephson frequency across the junction which matches the sum of the two mode frequencies, tunneling Cooper pairs excite photons in both modes simultaneously leading to far-from-equilibrium states. These states display highly non-classical features including strong anti-bunching, violation of Cauchy-Schwartz inequalitites, and number squeezing.  The regimes of low and high photon occupancies allow for analytical results which are supported by a full numerical treatment. The impact of asymmetries between the two modes is explored, revealing a pronounced enhancement of number squeezing when the modes are damped at different rates.
\end{abstract}
\maketitle
\section{Introduction}
It has long been known that the current flowing through a voltage-biased mesoscopic conductor can provide an extremely sensitive probe of its electromagnetic environment\,\cite{devoret,girvin,ingold,nazarov}. The current-voltage characteristics of a tunnel junction placed in series with a transmission line resonator is a particularly well-studied case\,\cite{devoret,girvin,holst}. The transmission line resonator contains a series of well-defined harmonic modes whose presence opens up inelastic current channels leading to characteristic features in the dc current flowing through the junction\,\cite{holst}.
The advent of high-Q superconducting resonators whose quantum state can be measured with great precision\,\cite{schoelkopf} together with the development of hybrid devices which couple non-metallic conductors to resonators\,\cite{qds1,frey}, has led to a renewed interest in the interaction between tunneling electrons or Cooper pairs and harmonic modes.
Whilst earlier experiments\,\cite{holst,wu} on mesoscopic conductors coupled to electromagnetic resonators focussed on how the harmonic modes affect the current in a regime where the modes themselves are close to thermal equilibrium, more recent experimental\,\cite{astafiev,hofheinz,petta,chen} and theoretical work\,\cite{rodrigues:07,marthaler:11,dqdl,berg,paduraiu:12,leppa:13,armour2013,gramich2013,kubala2014,schon2014,clerk2014} has begun to investigate how the current influences the resonator state and to explore the dynamics of systems where the resonator is far from thermal equilibrium.

For a Josephson junction which is biased with a sub-gap voltage, $V$, the relationship between the dc current and the energy pumped into the electromagnetic environment is particularly simple as all of the energy associated with a tunneling Cooper pair must be absorbed by the environment,\cite{hofheinz}. When the Josephson junction is placed in series with a transmission line resonator  a dc current is expected when the ac Josephson frequency $\omega_J=2eV/\hbar$ matches one or more of the mode frequencies in the transmission line. Experiments using low-Q resonators\,\cite{holst,hofheinz} have demonstrated that when the individual harmonic modes remain close to thermal equilibrium, they lead to well-defined peaks in the dc current whose heights and widths can be calculated using perturbation theory. In contrast, a high-Q resonator  can be excited to far-from-equilibrium states containing many photons\,\cite{chen} which are predicted to display intriguing non-classical features such as number squeezing\,\cite{armour2013,gramich2013}. This new field of Josephson photonics combines typical processes known from quantum optical set-ups with those known from charge transfer physics in highly versatile devices.

\begin{figure}[t]
\centering
{\includegraphics[width=6.0cm]{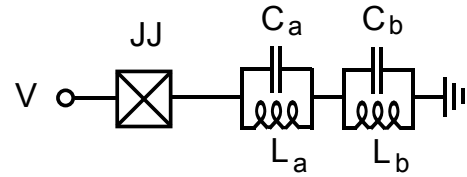}
}
\caption{Effective circuit model of the system. It consists of a Josephson junction (JJ) in series with two $LC$ oscillators, across which a voltage $V$ is applied. The two $LC$ oscillators are assumed to have different angular frequencies $\omega_a=(L_aC_a)^{-1/2}\neq\omega_b=(L_bC_b)^{-1/2}$.}
\label{fig:sketch}
\end{figure}

In this article we consider a voltage-biased superconducting junction whose ac Josephson frequency is tuned to excite {\em two} electromagnetic modes simultaneously (see Fig.\ \ref{fig:sketch}). Signatures of such processes have been observed in the dc current flowing through Josephson junctions coupled to low-Q resonators and can also be understood within a perturbative approach as the modes remain close to thermal equilibrium. While we address this domain as well, our main focus here lies in the regime  where the power transferred to the resonator modes is sufficient to drive them into far-from-equilibrium states while still displaying strong quantum properties. Note that the system we consider here differs from those used in recent experiments to produce photon pairs\,\cite{flurin,forgues} in that the energy comes from a dc voltage.

Starting from a simple model Hamiltonian which describes the effect of the Cooper pairs on the oscillators through a highly non-linear ac drive at the Josephson frequency, we use a rotating wave approximation to derive an effective time-independent Hamiltonian which we use to analyse the quantum dynamics of the oscillators. Although the full behavior of the system can only be uncovered by numerical solutions of the quantum master equation, we find that approximate  analytical descriptions are available in the two regimes of low and high photon occupancy. In the former one a perturbative treatment in the Josephson energy applies while in the latter explicit results are obtained by linearizing about the classical fixed points which provide a faithful description of the quantum dynamics when the zero-point fluctuations of the oscillators are small.

The excitation of the two oscillators shows a clear threshold as a function of the Cooper pair pumping rate.
Earlier work, which investigated the quantum dynamics of a single mode\,\cite{armour2013,gramich2013,kubala2014} driven by a voltage-biased Josephson junction, showed that non-classical features in the state of the oscillator such as number squeezing (sub-Poissonian photon statistics) occur very generally. For the two-mode system, we also find significant number-squeezing occurs in the states of the individual oscillators, especially in the above-threshold regime where the oscillators are strongly excited. Interestingly, when the damping rates of the oscillators are very unequal, the less-damped oscillator displays much stronger strong number squeezing than is ever found for a single-oscillator system. Provided that the quantum zero-point fluctuations are not too small, the number squeezing is strong enough to lead to negative regions in the Wigner function.

This work is organised as follows. We introduce our theoretical model in Sec.\ \ref{sec:model} and analyse its low photon limit in Sec.\ \ref{sec:lowphoton} and its semi-classical dynamics in Sec.\ \ref{sec:scd}. Sections \ref{sec:bth} and \ref{sec:ath} explore the quantum dynamics of the system in the below and above threshold regimes, respectively.
Finally, Sec.\ \ref{sec:conclude}  contains a discussion and the conclusions. The Appendix contains further details on some of the calculations described in the main text.

\section{Model System}
\label{sec:model}
We consider a system consisting of a Josephson junction in series with two $LC$ oscillators, $A$ and $B$ with angular frequencies $\omega_a$ and $\omega_b$ across which a voltage $V$ is applied (see Fig.\ \ref{fig:sketch}). The two oscillators could both be modes of a single superconducting resonator in which a Josephson junction is embedded between the ground plane and center conductor\,\cite{chen:11,blencowe:12,armour2013,chen} (See Ref.\ \onlinecite{armour2013} for a detailed derivation of the Hamiltonian for this case), but the system could also be realized using modes of two different electrical resonators\,\cite{holst}. The effective Hamiltonian of the system takes the form
\begin{eqnarray}
H&=&\hbar\omega_aa^{\dagger}a+\hbar\omega_bb^{\dagger}b \label{eq:modelh}\\
&&-E_J\cos\left[\omega_J t+\Delta_a(a+a^{\dagger})+\Delta_b(b+b^{\dagger})\right],\nonumber
\end{eqnarray}
where $E_J$ is the Josephson energy of the junction, $a$ and $b$ are the lowering operators of the oscillators with frequencies $\omega_a$ and $\omega_b$ respectively, and $\omega_J=2eV/\hbar$. The parameters $\Delta_{a(b)}$ quantify the strength of the zero-point fluctuations of the oscillators,
$\Delta_{a(b)}=(2e^2 Z_{a(b)}/\hbar)^{1/2}$ where $Z_{a(b)}=\sqrt{L_{a(b)}/C_{a(b)}}$ is the impedance.

Here we analyze the case where the system is operated close to the resonance that occurs when the voltage energy lost by a single Cooper pair traversing the circuit matches the energy required to simultaneously create one photon in each of the $LC$ oscillators, $\omega_J=2eV/\hbar=\omega_a+\omega_b$. We assume that the modes are not degenerate so that $\omega_a\neq\omega_b$. This means that the resonance at
$\omega_J=\omega_a+\omega_b$ does not compete with processes in which two photons are absorbed by just one of the modes.

We examine the behavior of the system as a function of the Josephson energy which describes the strength of the Cooper pair tunneling. The value of $E_J$ can be thought of like a pumping rate for the oscillators: as it is increased the oscillators will be more strongly driven, become more strongly excited and behave more non-linearly. In practice $E_J$ can be varied in an effective single-junction by forming two junctions in parallel and applying a tunable flux in the SQUID loop that they form\,\cite{nakamura,armour2013}.

The strengths of the quantum fluctuations parameterised by $\Delta_{a}$, $\Delta_{b}$, also play a very interesting role in determining the dynamics of the system and we will examine how the behavior is modified when they are varied. For systems where a Josephson junction is embedded in a superconducting resonator designed to have a very high-Q the quantum fluctuations will typically be very small $\Delta_{a(b)}\ll 1$ . However, significantly stronger quantum fluctuations have very recently been engineered in low-Q resonators coupled to tunnel junctions\,\cite{alti} and it may be possible to combine stronger quantum fluctuations with higher Q values in the future.

\subsection{Rotating wave approximation}
The explicit time-dependence in the Hamiltonian complicates the analysis of the corresponding dynamics significantly. However, close to the resonance we are interested in, $\omega_J\simeq\omega_a+\omega_b$, only some of the terms will play an important role and these can be picked out by a rotating wave approximation (RWA).

 We proceed following the approach in Refs.\ \onlinecite{armour2013,gramich2013,kubala2014}. We move to a rotating frame, applying a unitary transformation of the form $U(t)={\rm e}^{i\tilde{\omega}_aa^{\dagger}at}{\rm e}^{i\tilde{\omega}_bb^{\dagger}bt}$ where we define $\tilde{\omega}_a+\tilde{\omega}_b=\omega_J$, and make a RWA in which we neglect all of the rapidly oscillating terms in the rotating frame. The resulting effective Hamiltonian takes the form,
\begin{eqnarray}
H_{\rm RWA}&=&\hbar\delta^{(a)}a^{\dagger}a+\hbar\delta^{(b)}b^{\dagger}b \label{eq:hrwa}\\
&&+\frac{\tilde{E}_J}{2}:\frac{J_1(2\Delta_a\sqrt{a^{\dagger}a})
J_1(2\Delta_b\sqrt{b^{\dagger}b})}{\sqrt{a^{\dagger}a}\sqrt{b^{\dagger}b}}\left(a^{\dagger}b^{\dagger}+ab\right):, \nonumber
\end{eqnarray}
where the colons imply normal ordering of the operators, $\delta^{(i)}=\omega_i-\tilde{\omega_i}$ for $i=a,b$ and $\tilde{E_J}=E_J{\rm e}^{-(\Delta_a^2+\Delta_b^2)/2}$. For sufficiently low photon numbers (such that $2\Delta_a\sqrt{\langle a^{\dagger}a\rangle}, 2\Delta_b\sqrt{\langle b^{\dagger}b\rangle}\ll 1$) we can expand the Bessel functions in Eq.\ \eqref{eq:hrwa} to lowest order. In this limit the system reduces to a non-degenerate parametric amplifier\cite{milburn}
\begin{equation}
H^{(0)}_{\rm RWA}=\hbar\delta^{(a)}a^{\dagger}a+\hbar\delta^{(b)}b^{\dagger}b+\frac{\tilde{E}_J\Delta_a\Delta_b}{2}\left(a^{\dagger}b^{\dagger}+ab\right). \label{eq:ndpa}
\end{equation}

\subsection{Quantum master equation}

\label{sec:qme}
The two oscillators are assumed to be weakly damped at rates $\gamma_a$ and $\gamma_b$ which in general will not be the same. We therefore assume that the quantum master equation of the system takes the standard quantum optical form in the $T=0$ limit\,\cite{milburn}
\begin{eqnarray}
\frac{d{\rho}}{d\tau}&=&-{i}[\tilde{H}_{\rm RWA},\rho]+\frac{r}{2}\left(2a\rho a^{\dagger}-a^{\dagger}a\rho-\rho a^{\dagger}a\right)\nonumber\\
&&+\frac{1}{2r}\left(2b\rho b^{\dagger}-b^{\dagger}b\rho-\rho b^{\dagger}b\right), \label{eq:qme}
\end{eqnarray}
where we adopt dimensionless units of time $\tau=t\sqrt{\gamma_a\gamma_b}$, $r=\sqrt{\gamma_a/\gamma_b}$ and $\tilde{H}_{\rm RWA}=H_{\rm RWA}/(\hbar\sqrt{\gamma_a\gamma_b})$.

In an actual experimental realization of the JJ-oscillators system in Fig.\ \ref{fig:sketch}, the damping of the oscillators (due to photon decay from the resonators) is not the only source of dissipation. Indeed, the existence and impact of local voltage fluctuations at the JJ can be seen in the broadening of the spectrum of emitted microwave radiation\,\cite{hofheinz,gramich2013}. The existence of such fluctuations necessitates including explicitly an extra degree of freedom for the number of Cooper pairs $N$ transported across the junction in the model.  In the effective Hamiltonian, Eq.\ \eqref{eq:hrwa}, the $\left(a^{\dagger}b^{\dagger}+ab\right)$ term then gets replaced by $\left(e^{i\eta}\, a^{\dagger}b^{\dagger}+e^{-i\eta}\,ab\right)$, where $e^{\pm i\eta}=\sum_N |N\rangle \langle N\pm1|$.
Local voltage fluctuations are included by an additional dissipator in (\ref{eq:qme}) which in the simplest version takes the form  ${\cal L}[N,\rho]=r_J(2 {N}\rho\, {N}-{N}^2\, \rho-\rho\, {N}^2)$ with $r_J=\gamma_J/\sqrt{ \gamma_a \gamma_b}$. Ref.~\onlinecite{gramich2013}  describes, how to treat the corresponding quantum master equation in the extended JJ-resonator space.

However, it turns out  that only certain observables sensitively depend on the strength of these fluctuations, characterized by $\gamma_J$, for example the spectral broadening. For other observables, such as the photon occupation and photonic correlation functions that are of relevance for this work, the impact of local voltage fluctuations is basically negligible since experimentally one typically has $\gamma_J\ll \gamma_{a,b}$ (see for example Ref.~\onlinecite{hofheinz}). Then, formally, the Hamiltonian (\ref{eq:hrwa}) is regained by putting $\gamma_J=0$ so that the phase operators ${\rm e}^{\pm i\eta}$  simply appear as phase factors which can be removed via the gauge transformation ${\rm e}^{i\eta/2} a^\dagger, {\rm e}^{i\eta/2} b^\dagger \to a^\dagger, b^\dagger$. Note that this reflects a phase invariance of the RWA Hamiltonian (\ref{eq:hrwa}).

 \subsection{Relevant observables}

 The basic structure of the RWA Hamiltonian [Eq.\ \eqref{eq:hrwa}] in which photons are always created (or destroyed) jointly in the two oscillators and the linear damping that we assumed in formulating the master equation lead to a simple connection between the occupation numbers of the two modes $n_{a(b)}=\langle a^\dagger a (b^\dagger b)\rangle$ and the average dc current, $I_{\rm dc}$, flowing through the junction that can be obtained from an energy balance argument without the need to work with a current operator. Since each Cooper-pair that contributes to the dc current must create exactly one additional photon in {\rm each} of the oscillators, the requirement that the energy gain and loss rates balance tells us that
\begin{equation}
\frac{I_{\rm dc}}{2e}=\gamma_a n_a=\gamma_b n_b, \label{eq:balance}
\end{equation}
where in this case we have returned to dimension-full units.

The quantum nature of the photonic states in the oscillators is captured by photon correlation functions such as
\begin{equation}\label{g2}
g_{aa(bb)}^{(2)}(0)= \frac{\langle [a^\dagger a ( b^\dagger b)]^2\rangle -n_{a(b)}}{n_{a(b)}^2}\, , \, g_{ab}^{(2)}(0)=\frac{\langle a^\dagger a b^\dagger b\rangle}{n_a n_b}
\end{equation}
and the Fano factors
\begin{equation}\label{fano}
F_{a(b)}=\frac{\langle[a^{\dagger}a (b^{\dagger}b)]^2\rangle-n_{a(b)}^2}{n_{a(b)}}\, .
\end{equation}
Whilst these two types of correlation functions are closely related to each other, they are nevertheless useful to characterize the photonic states in opposite regimes of parameter space. In the regime of weak driving and low photon occupation deviations from the case of a driven harmonic oscillator are best seen in the $g^{(2)}$ functions. Namely, with increasing driving amplitude $\tilde{E}_J$, the photon distributions for the number states in the cavities evolve from  Poissonian distributions with  almost empty cavities towards  distributions peaked
 around finite mean occupations $n_a, n_b$. In this case the $g^{(2)}(0)$ functions (\ref{g2}) sensitively indicate deviations from the linear regime  $g^{(2)}_{aa(bb)}(0)\equiv 1$ with $g_{ab}^{(2)}(0)\neq 0$ capturing growing cavity-cavity correlations.
In the opposite regime of strong driving, nonlinearities may substantially influence the widths of the peaks for photon occupations (energy fluctuations) as properly measured in the Fano-factors (\ref{fano}).

In the following, we will first focus on the regime of low photon occupancy, where charge transfer through the Josephson junction occurs sequentially (Coulomb blockade regime) and analytical results can be obtained via a perturbative treatment in the drive amplitude ${E}_J$. In the opposite domain of large photons numbers in the cavities, a semi-classical type of approximation applies, though, with substantial quantum fluctuations. In both domains, the magnitude of the parameters $\Delta_a, \Delta_b$, i.e.\ the strength of the ground state fluctuations of the respective cavities, plays a decisive role: they rule the impact of nonlinearities in the former regime and control the impact of quantum effects in the latter one. Corresponding findings will be supported by numerical solutions of the stationary states according to (\ref{eq:hrwa}) and (\ref{eq:qme}).

\section{Few-Photon limit}
\label{sec:lowphoton}
The physics of the system described by the Hamiltonian \eqref{eq:hrwa} and the master equation \eqref{eq:qme}
is at its simplest when it is driven so weakly that excitations in the resonators will relax to equilibrium well before a new excitation occurs. In that regime, very few photons, $n_{a/b} \ll 1$, reside in the resonators on average. Transport across the junction in turn is in the (dynamical) Coulomb-blockade regime, where subsequent Cooper-pair tunneling events occur uncorrelated with some tunneling rate. While the charge flows uncorrelated, the photons exhibit correlations already at the weakest driving.

Now, for the present set-up one derives from the full RWA Hamiltonian
\begin{equation}\label{fulln}
n_{a}=\frac{i E_J}{2 E_J^c\, r}\langle : \left(a b-a^\dagger b^\dagger\right) \frac{J_1(2\Delta_a\sqrt{a^\dagger a})}{\Delta_a\sqrt{a^\dagger a}}\frac{J_1(2\Delta_a\sqrt{b^\dagger b})}{\Delta_b\sqrt{b^\dagger b}}:\rangle
\end{equation}
with $E_J^c=(\hbar\sqrt{\gamma_a\gamma_b}/\Delta_a \Delta_b) {\rm e}^{(\Delta_a^2+\Delta_b^2)/2}$ and where $n_b$ follows by replacing $r\to 1/r$. In the lowest order in the driving strength this reduces to
\begin{equation}\label{eq:fewphot_n}
n_{a}^{(0)} = \frac{1}{4}\left(\frac{{E}_J}{E_J^c}\right)^2  \frac{
1+r^2
}{r^2 (\delta^{(a)}+\delta^{(b)})^2 +(1+r^2)^2/4}\, ,
\end{equation}
with the superscript  indicating the  leading order in $E_J^2$ and with $n_b^{(0)}$ again following from $r\to 1/r$.
\begin{figure}[h]
\centering
\includegraphics[width=8cm]{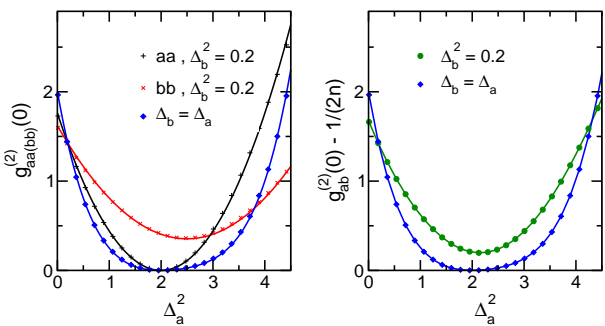}
\caption{(Color online) Autocorrelations $g_{aa(bb)}^{(2)}(0)$ (left) and cross-correlations
$g_{ab}^{(2)}(0)$ (right) of the two modes vary with the strength of zero-point
fluctuations $\Delta_{a(b)}$ in the two oscillators. For weak driving, $E_J = 0.2E_J^c$, the autocorrelations (symbols)
are given by (\ref{eq:gaa}) (lines) when $\Delta_a$, or simultaneously $\Delta_a$ and
$\Delta_b$ are tuned. The reduced cross-correlations
$g_{ab}^{(2)}(0) - 1/(2n)$ (lines) obey the general relation (\ref{eq:gab}) with the mean of the autocorrelations
$[g_{aa}^{(2)}(0)+g_{bb}^{(2)}(0)]/2$ depicted as symbols for the case of symmetric damping $r=1$.}
\label{fig:g2kappa}
\end{figure}

For the correlations we focus on the symmetric case $\gamma_a=\gamma_b$ at resonance so that $n_a=n_b=n$.
Then, based on the full Hamiltonian (\ref{eq:hrwa}) one can show the general relation
\begin{equation}
\langle a^\dagger a b^\dagger b\rangle = \frac{n}{2} +\frac{n^2}{2}\left[g_{aa}^{(2)}(0)+g_{bb}^{(2)}(0)\right]\,
\end{equation}
which implies
\begin{equation}\label{eq:gab}
g_{ab}^{(2)}(0)=\frac{1}{2 n} +\frac{1}{2}\left[g_{aa}^{(2)}(0)+g_{bb}^{(2)}(0)\right]\,
\end{equation}
with $n$ as given in (\ref{fulln}). Now, working to order $E_J^4$, one finds
\begin{equation}
g_{aa(bb)}^{(2)}(0)=2\left(1-\frac{\Delta_{a(b)}^2}{2}\right)^2 \left(1-\frac{5}{8}\Delta_{b(a)}^2+\frac{\Delta_{b(a)}^4}{8}\right)\, . \label{eq:gaa}
\end{equation}

 Two types of correlations are encoded in the above $g^{(2)}(0)$ functions. The most obvious ones stem from the common excitation process of photons in the two resonators. They are therefore already present in the parametric amplifier limit of the Hamiltonian \eqref{eq:ndpa} and well understood for that case, see e.g. Ref.\ \onlinecite{graham}. A convenient tool to characterise them is the noise reduction factor\,\cite{forgues} ${\rm NRF}=[\langle (a^\dagger a-b^\dagger b)^2\rangle-(n_a-n_b)^2]/(n_a+n_b)$ which in the symmetric situation $\gamma_a=\gamma_b$ takes the form
\begin{equation}
{\rm NRF}=\frac{n}{2}\left[g_{aa}^{(2)}(0)+g_{bb}^{(2)}(0)-2g_{ab}^{(2)}(0)\right]+1\, .
\end{equation}
However, the perfect correlation of the excitation process  leads to perfectly correlated occupations in the oscillators with a noise reduction factor ${\rm NRF}=0$ {\em only} for the undamped case $\gamma_a=\gamma_b=0$. For any finite photon lifetimes in the cavities, the decay out of the two cavities occurs uncorrelated which according to (\ref{eq:gab}) always implies in the stationary state and for the symmetric situation  ${\rm NRF}=1/2$.

Further correlations in the light field are caused by the back-action of the resonator occupations on the photon creation processes. Generally speaking, the existence of photonic excitations in the resonators can either increase the probability of further excitations, similar to a stimulated emission effect, or it can hinder further excitations. Formally, these effects are encoded in the transition matrix elements of the RWA-Hamiltonian \eqref{eq:hrwa} between neighboring oscillator states, where the nonlinearities of the Bessel functions enter.
If charge quantization of the Cooper-pair current is significant, the parameters $\Delta_{a/b}$ become large, so that the nonlinearities already appear at the few photon level. For the case of a single resonator, it was shown in Ref.~\onlinecite{gramich2013} that $\Delta^2=2$ can completely suppress transitions to higher occupations and reduces the resonator effectively to a two-level system, thus operating as a perfect single photon source. The behavior of the correlation functions in the two-mode case is shown in Fig.\ \ref{fig:g2kappa}. While a non-zero $g_{aa}^{(2)}(0)$ requires oscillator $a$ to be populated up to the second excited state by two successive photons, this need not be the case for oscillator $b$ as it can relax before the second photon arrives. Consequently, as seen in (\ref{eq:gaa}), $g_{aa}^{(2)}(0)=0$ at $\Delta_a^2 =2$,  but not at $\Delta_b^2=2$.

 The general result (\ref{eq:gab}) also reveals that the classical Cauchy-Schwartz inequality for photon intensities is {\em always} violated in the quantum case, i.e.,
 \begin{equation}
 \sqrt{g_{aa}^{(2)}(0)\,  g_{bb}^{(2)}(0)} \leq g_{ab}^{(2)}(0)\, .
 \end{equation}
 Namely, introducing the parameter $\epsilon=g_{bb}^{(2)}(0)/g_{aa}^{(2)}(0)$ the violation of the inequality requires $[- g_{aa}^{(2)}(0)] (1-\sqrt{\epsilon})^2 \leq 1/n$ which always applies since $g_{aa}^{(2)}(0), n\geq 0$.
 Accordingly, emission of photons from the cavities occurs in a correlated way for all driving strengths and photon occupations. In the next Section we ascribe to the individual photon states in the cavities respective amplitudes (energies) and phases. One then sees that these states are correlated through their {\em phases} due to the simultaneous creation process in  the transfer of a single Cooper pair.

\section{Semi-Classical Dynamics}\label{sec:scd}

For large photon occupancy in the cavities a semiclassical type of approximation applies. The simplest semi-classical description of the dynamics of the system is obtained from the equations of motion for $\langle a \rangle$ and $\langle b\rangle$ which follow from Eq.\ \eqref{eq:qme}, making the replacements $\langle a \rangle=\alpha$, $\langle b\rangle= \beta$ and treating expectation values of products of operators as products of expectation values. Hence we find
 \begin{eqnarray}
 \dot{\alpha}&=&-\left(i\tilde{\delta}^{(a)}+\frac{r}{2}\right)\alpha+\frac{i{E}_J}{2\Delta_b{E}_J^c}J_1(2\Delta_b|\beta|)\times\nonumber\\
 &&\left[J_2(2\Delta_a|\alpha|)\frac{\alpha^2\beta}{|\alpha|^2|\beta|}-J_0(2\Delta_a|\alpha|)\frac{\beta^*}{|\beta|}\right] \label{eq:alpha1}\\
 \dot{\beta}&=&-\left(i\tilde{\delta}^{(b)}+\frac{1}{2r}\right)\beta+\frac{i{E}_J}{2\Delta_a{E}_J^c}J_1(2\Delta_a|\alpha|)\times\nonumber\\
 &&\left[J_2(2\Delta_b|\beta|)\frac{\beta^2\alpha}{|\beta|^2|\alpha|}-J_0(2\Delta_b|\beta|)\frac{\alpha^*}{|\alpha|}\right] \label{eq:beta1},
 \end{eqnarray}
where $\tilde{\delta}^{(a,b)}={\delta}^{(a,b)}/\sqrt{\gamma_a\gamma_b}$. Obtained in this way, the factors of ${\rm e}^{(\Delta_a^2+\Delta_b^2)/2}$ embodied in ${E}_J^c$ that appear in these equations are accidental: they would not be present if we had instead chosen to use a symmetric ordering for the operators when deriving the Hamiltonian. However,  Eqs.\ \eqref{eq:alpha1} and \eqref{eq:beta1} would also arise from a simple-minded ansatz in which we assumed that the density operator of the system is just a product of the coherent states $\rho(t)=|\alpha(t)\rangle\langle \alpha(t)|\otimes|\beta(t)\rangle\langle \beta(t)|$, in this approximation the factors of ${\rm e}^{(\Delta_a^2+\Delta_b^2)/2}$ would arise naturally.

Using amplitude-phase coordinates for the two oscillators, $\alpha=A{\rm e}^{-i\phi_a}$ and $\beta=B{\rm e}^{-i\phi_b}$, and introducing the total and relative phase variables $\xi^{\pm}=\phi_a\pm\phi_b$, Eqs.\ \eqref{eq:alpha1} and \eqref{eq:beta1} take the form
\begin{eqnarray}
\dot{A}&=&-\frac{r}{2}A+\frac{{E}_J}{{E}_J^c}\frac{J_1(2\Delta_bB) J_1(2\Delta_a A)}{2\Delta_a\Delta_b A}\sin(\xi_+)\label{eq:ampa}\\
\dot{B}&=&-\frac{1}{2r}B+\frac{{E}_J}{{E}_J^c}\frac{J_1(2\Delta_aA)J_1(2\Delta_bB) }{2\Delta_a\Delta_b B}\sin(\xi^+)\label{eq:ampb} \\
\dot{\xi}^+&=&\delta^{(+)}+F_+(A,B)\cos\xi^+ \label{eq:psum}\\
\dot{\xi}^-&=&\delta^{(-)}+F_-(A,B)\cos\xi^+,\label{eq:pdiff}
\end{eqnarray}
where we used the Bessel function identity, $J_2(z)+J_0(z)=2J_1(z)/z$, and have defined $\delta^{(\pm)}=\tilde{\delta}^{(a)}\pm\tilde{\delta}^{(b)}$. Further,
\begin{eqnarray}
F_{\pm}(A,B)&=&\frac{E_J}{2E_J^c}\left(\frac{J_1(2\Delta_b B)}{\Delta_bA}\left[J_0(2\Delta_aA)-J_2(2\Delta_aA)\right]\right.\nonumber\\
&\pm&\left.\frac{J_1(2\Delta_a A)}{\Delta_a B}\left[J_0(2\Delta_bB)-J_2(2\Delta_bB)\right]\right)
\end{eqnarray}
with the property $F_+(-A,B)=F_-(A,B)$ and $F_+(A,-B)=-F_-(A,B)$.
The behavior of the system is determined by the fixed points of the amplitudes $A_0,B_0$ and the total phase $\xi^+_0$. Since the relative phase does not appear on the righthand side of any of these equations its fixed point value is arbitrary. For simplicity, we concentrate on the on-resonance case $\delta^{(a)}=\delta^{(b)}=0$ in our analysis.

The amplitude equations lead to the fixed point conditions $A_0=B_0=0$ or
\begin{eqnarray}
\sin\xi^+&=&\frac{r\Delta_a\Delta_bE_J^cA_0^2}{{E}_J J_1(2\Delta_bB_0)J_1(2\Delta_aA_0)}\\
&=&\frac{\Delta_a\Delta_bE_J^cB_0^2}{r{E}_J J_1(2\Delta_bB_0)J_1(2\Delta_aA_0)}\,. \label{eq:ampc}
\end{eqnarray}
The second equality in Eq.\ \ref{eq:ampc} leads to the energy balance condition $B_0=rA_0$. From the equation for $\xi^+$, we see that fixed points arise when either $\cos\xi^+_0=0$ or
\begin{equation}
F_+(A_0,B_0)=0. \label{eq:fb2}
\end{equation}
This latter condition is independent of $E_J$ and hence leads to a locking of the amplitudes at particular values as a function of $E_J$, something which is an important characteristic of the dynamics in the single-oscillator system\,\cite{armour2013}.
For symmetric oscillators ($r=1$ and $\Delta_a=\Delta_b$) $F_+=0$ implies $J'_1(z)=0$ with $z=2\Delta_a A_0=2\Delta_b B_0$ which has a first solution at $z=1.841$\,\cite{armour2013}.

Thus we identify three possible fixed points for the system: a zero-amplitude one, one given by the conditions $\cos\xi^+=0$ and (from Eq.\ \eqref{eq:ampc})
\begin{equation}
\frac{r A_0^2\Delta_a\Delta_bE_J^c}{{E_J}J_1(2\Delta_brA_0)J_1(2\Delta_a A_0)}=\pm 1, \label{eq:abx}
\end{equation}
and a third solution for which the amplitudes lock to values where Eq.\ \eqref{eq:fb2} is satisfied (together with the condition $B=rA$)  and the total phase is be given by Eq.\ \eqref{eq:ampc}.

We can look for small amplitude solutions to Eq.\ \ref{eq:abx} ($\Delta_b rA_0,\Delta_a A_0\ll 1$) by expanding the Bessel functions and retaining the lowest order terms in $A_0$,
\begin{equation}
A_0=\sqrt{2\frac{\left(1-\frac{E_J^c}{E_J}\right)}{\Delta_b^2r^2+\Delta_a^2}}.
\end{equation}
Thus we see that a non-zero amplitude solution only exists for $E_J>E_J^c$. Thus $E_J^c$ has a simple physical interpretation: it is the value of $E_J$ at which the oscillators reach the threshold for non-zero amplitude oscillations.

Taking into account the stability of the fixed points, we find that as $E_J$ is increased from zero the amplitudes remain zero until the system reaches  threshold at $E_J=E_J^c$, after which the amplitudes grow smoothly according to Eq.\ \eqref{eq:abx} with the global phase locked to $\xi^+_0=\pi/2$. For a sufficiently large $E_J$, which we define as $E_J^{c2}$, a bifurcation occurs as the amplitudes become large enough to satisfy Eq.\ \eqref{eq:fb2} and the amplitudes then lock, becoming independent of $E_J$.


In the next two sections we will examine the quantum dynamics of the system in the below and above threshold regimes.

\section{Sub-threshold dynamics}

\label{sec:bth}


In the sub-threshold regime ($E_J< E_J^c$) the semi-classical fixed points have zero amplitude ($A=B=0$). In this case we can gain some insight into the behavior of the system by approximating the Hamiltonian of the system by its lowest order terms, i.e.\  setting $H_{\rm RWA}=H^{(0)}_{\rm RWA}$ [see Eq.\ \ref{eq:ndpa}], an approach which is equivalent to analysing the linear fluctuations about the semi-classical fixed points.

When this approximation is made the Hamiltonian is quadratic and the equations of motion for the moments take a rather simple form. Solving these equations, we find in the steady-state
\begin{eqnarray}
 n_a &=& r^{-2}n_b=\frac{\left(\frac{E_J}{E_J^c}\right)^2}{\left[1+r^2\right]\left[1-\left(\frac{E_J}{E_J^c}\right)^2\right]} \label{eq:aven}\\
\langle  a b\rangle &=&-i\left(\frac{r}{r^2+1}\right)\frac{\left(\frac{E_J}{E_J^c}\right)}{1-\left(\frac{E_J}{E_J^c}\right)^2\label{eq:cross} }\\
\langle a\rangle&=&\langle b\rangle=\langle a b^{\dagger}\rangle=0\, .
\end{eqnarray}
We note in passing that the result for $n_a$ reduces to the one derived in (\ref{eq:fewphot_n}) in leading order in $E_J/E_J^c$.

Simplified in this way, the linearized description leads to a Gaussian steady-state Wigner function which takes the form\cite{agarwal:90,agarwal:book}
\begin{equation}
W_{a,b}(\alpha,\beta)=\frac{{\rm e}^{-\left[(n_b+1/2)|\alpha|^2+(n_a+1/2)|\beta|^2+\mu\alpha\beta+\mu^*\alpha^*\beta^*\right]/C}}{\pi^2 C} \label{eq:wig}
\end{equation}
where $C=\left[(n_a+1/2)(n_b+1/2)-|\mu|^2\right]$ and $\mu^*=-\langle ab\rangle $. This is a mixed state which combines two-mode squeezing and thermal-like fluctuations\cite{agarwal:book}.
The Wigner function of the individual oscillators is obtained by integrating over the phase space of the other one leading in either case to a thermal distribution. Thus for oscillator a, for example, we have
\begin{equation}
W_a(\alpha)=\frac{1}{\pi(n_a+1/2)}\exp\left[-\frac{|\alpha|^2}{(n_a+1/2)}\right]. \label{eq:wa}
\end{equation}

\begin{figure}[h]
\centering
{\includegraphics[width=7.0cm]{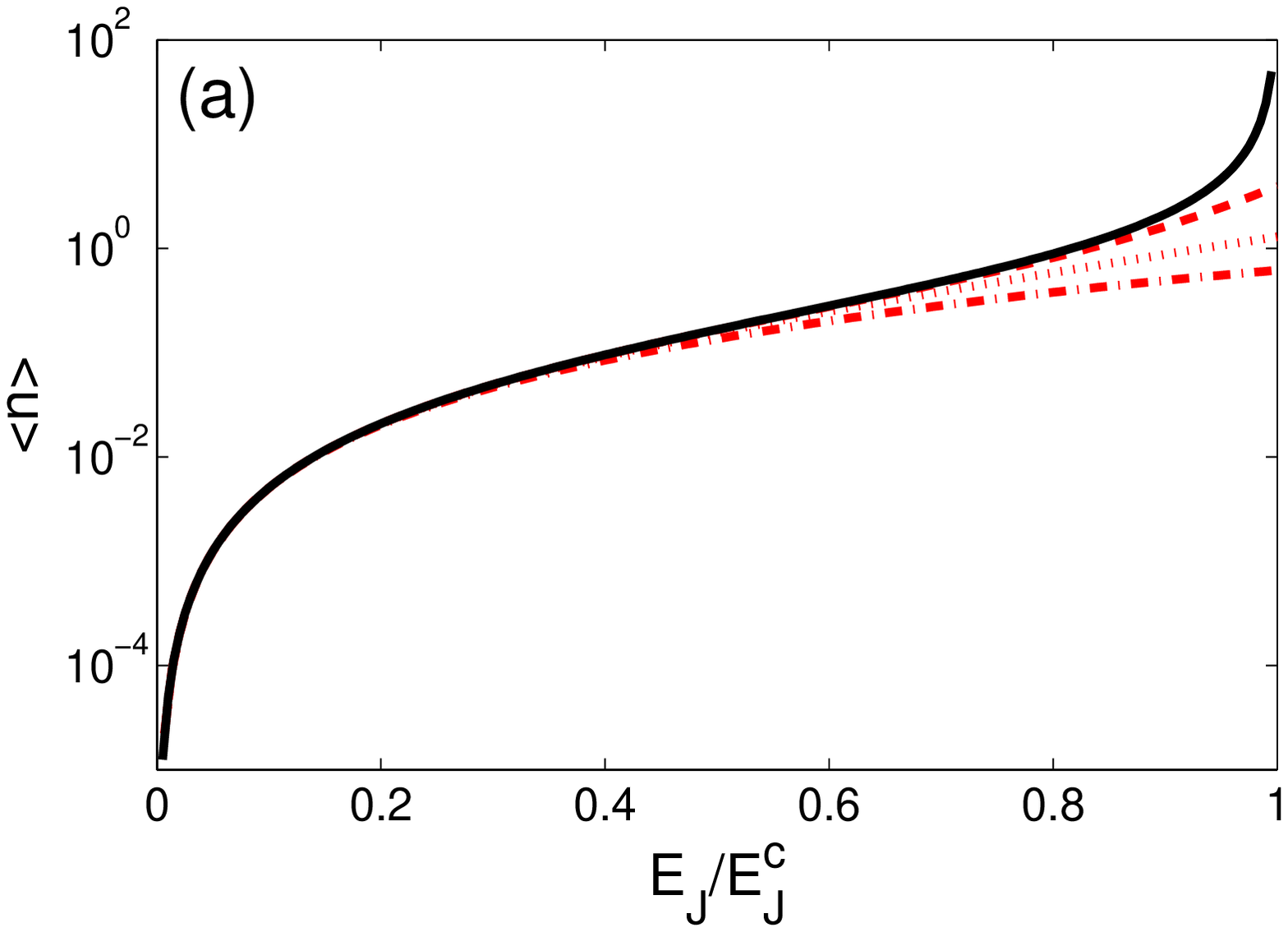}
\includegraphics[width=7.0cm]{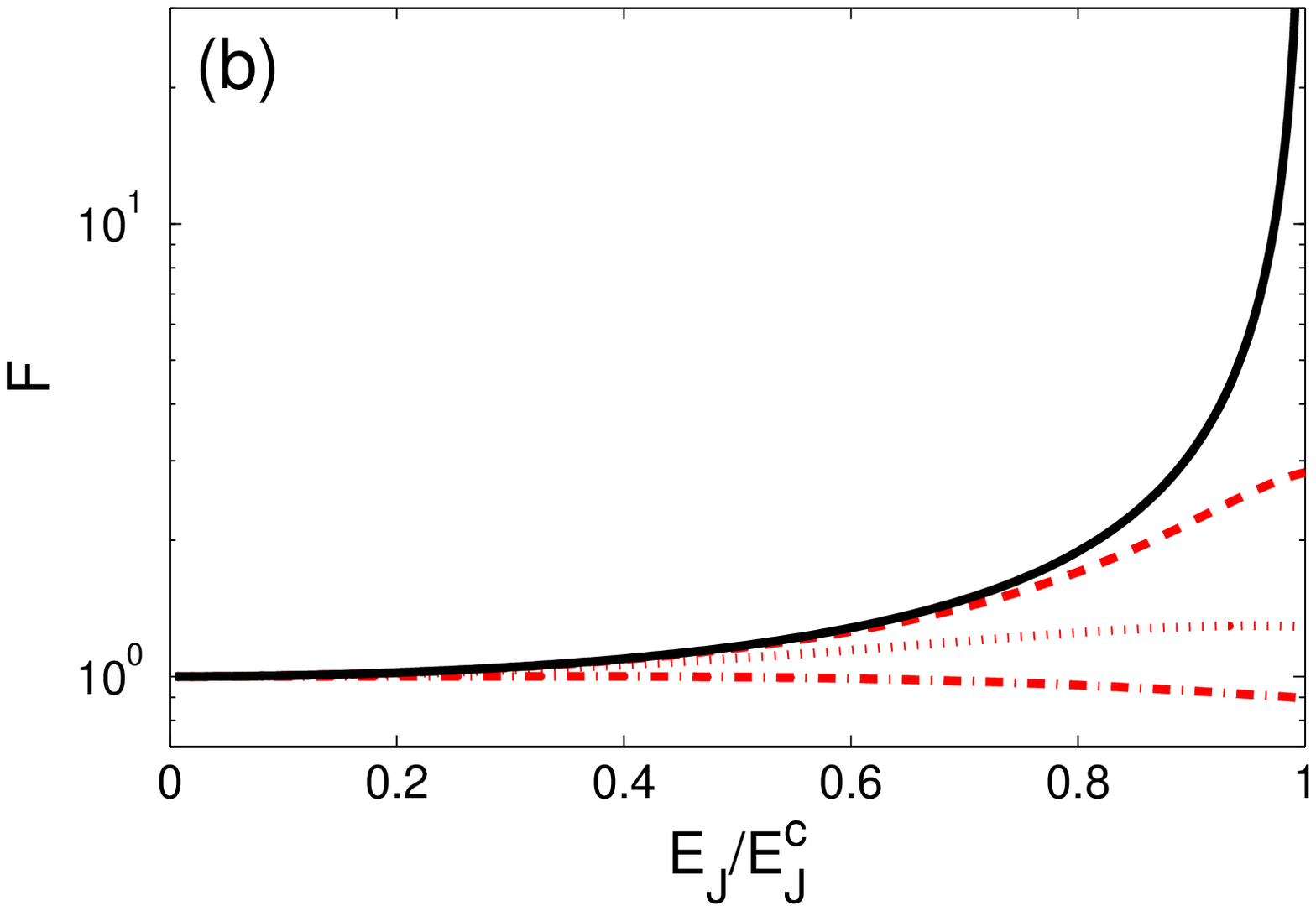}
}
\caption{(Color online) (a) Average occupation, $\langle n\rangle=n_a=n_b$, (b) Fano factor, $F=F_a=F_b$,
 as a function of $E_J/E_J^c$ for symmetric oscillators. The full curves are the linearized results and the other curves are for $\Delta=0.1$ (dashed curves), $\Delta=0.3$ (dotted curves) and $\Delta=0.6$ (dash-dotted curves).}
\label{fig:delta}
\end{figure}


The full behavior of the average energy of oscillator $a$, $n_a$,  obtained by solving the master equation numerically\cite{qutip}, is shown in Fig.\ \ref{fig:delta} for symmetric oscillators ($r=1$, $\Delta=\Delta_a=\Delta_b$). The divergence in $n_a$ which the linearized analysis predicts for $E_J\rightarrow E_J^c$ [Eq.\ \eqref{eq:aven}] never occurs in the full quantum problem as higher order terms in the RWA Hamiltonian always saturate the energy gain. As $\Delta$ is increased the saturation occurs at progressively lower values of the photon number whilst the range of $E_J/E_J^c$ values for which the linearized calculation is accurate becomes smaller and smaller.

The fluctuations in the energy of the oscillators, described by the Fano factors $F_{a(b)}$ (\ref{fano}) change rather more dramatically with $\Delta_a$. The thermal Wigner function obtained from the linearized calculation [Eq.\ \eqref{eq:wa}] predicts the simple relationship between Fano factor and photon number associated with thermal states, $F_{a(b)}=n_{a(b)}+1$,  leading to growth in $F_{a(b)}$ as $E_J/E_J^c$ increases and again there is a divergence at threshold. For small values of $\Delta$, the full quantum dynamics follows a similar pattern though with saturation in $F_{a(b)}$ at the threshold leading to a peak rather than a divergence. In contrast, for larger $\Delta$ values the behavior is completely different: the value of $F_{a(b)}$ {\em drops} monotonically as $E_J/E_J^c$ is increased and its behavior contains no signature of the threshold at $E_J^c$.

 The change in the behavior of $F_a$ as $\Delta_a$ is increased is reminiscent of quantum optical systems like the laser\,\cite{carmichael}, which in the `thermodynamic' limit of weak atom-photon couplings display clear thresholds (accompanied by a signature peak in the Fano factor) whose properties can be understood in terms of an analogy with classical phase transitions, but which for sufficiently strong couplings behave quite differently without clear signatures of a threshold\,\cite{carmichael,ionlaser}.

\section{Dynamics above threshold}
\label{sec:ath}

Above threshold the oscillators become strongly excited though this does not mean that their states become classical. As in the case of the single-oscillator system\,\cite{armour2013}, strong number squeezing (marked by a Fano factor below unity) occurs even at large average occupation numbers. As in the sub-threshold regime, the behavior of the system in the limit of very small zero-point fluctuations,  $\Delta_a,\Delta_b\ll 1$, can be captured within an approximate description which linearizes about the semi-classical fixed points of the system, but for larger zero-point fluctuations numerical solution of the quantum master equation becomes essential. We start by exploring the general properties of the steady-states of the individual oscillators in the above-threshold regime for symmetric oscillators and the role played by the size of the zero-point fluctuations before going on to examine how asymmetry alters the behavior.


\subsection{Symmetric Oscillators}
\begin{figure}[t]
\centering{\includegraphics[width=7.0cm]{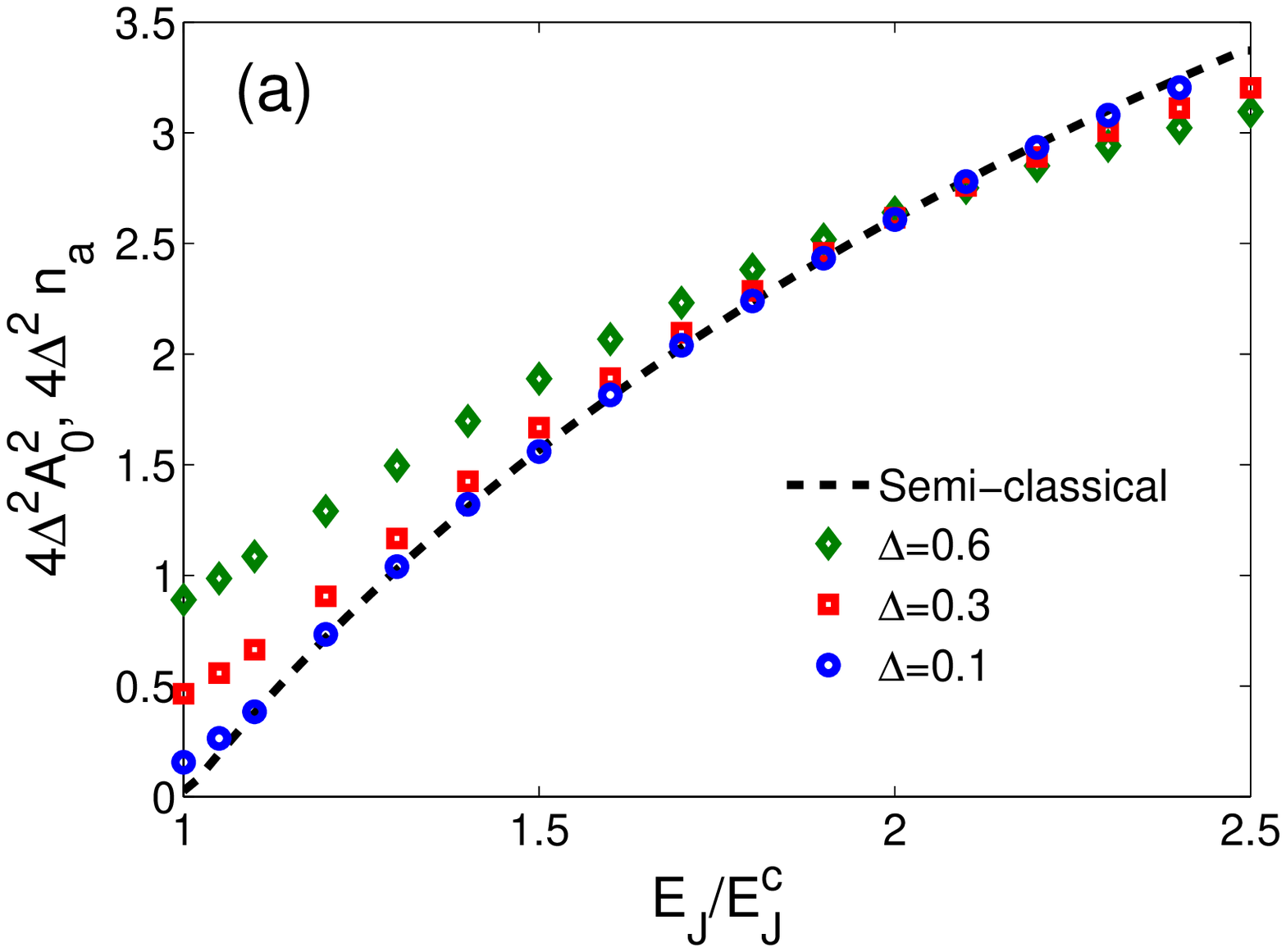}
\includegraphics[width=7.0cm]{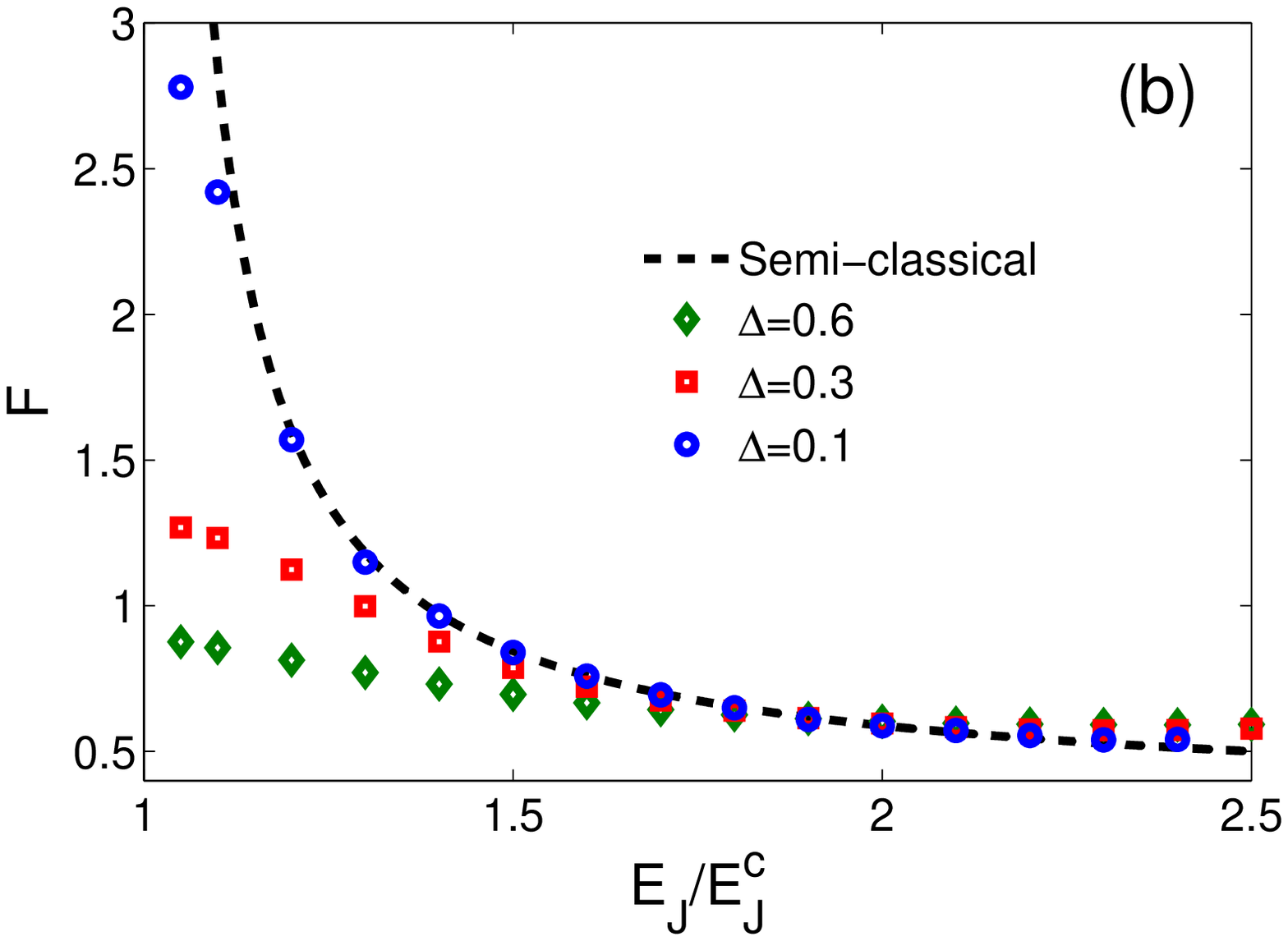}}
\caption{(Color online) Comparison of the oscillator occupation numbers (a) and Fano factor (b) obtained from numerical solution of the quantum master equation for $\Delta=0.1$, $0.3$ and $0.6$ with corresponding semi-classical calculations over the range $E_J^c<E_J<E_J^{c2}=2.5E_J^c$. In (a) both the semi-classical oscillator energy, $A_0^2$, and occupation number, $n_a$ are scaled by $4\Delta^2$. }
\label{fig:scaled}
\end{figure}

For symmetric oscillators ($r=1$, $\Delta_a=\Delta_b=\Delta$) the steady-state properties of the two oscillators must be the same and there is a very simple scaling to the semi-classical fixed point amplitudes obtained in Sec.\ \ref{sec:scd}:  the value of $2\Delta_aA_0$ is a function of just $E_J/E_J^c$, see \eqref{eq:ampc}. This scaling provides a convenient way of comparing the average oscillator occupation $n=n_a=n_b$ (obtained by solving the master equation numerically) for different values of $\Delta$  with the semi-classical prediction, as shown in Fig.\ \ref{fig:scaled}a. We solved the master equation using standard numerical methods\,\cite{qutip}; for smaller values of $\Delta$ we carried out quantum trajectory simulations, whilst for larger $\Delta$ we were able to solve for the steady-state of the master equation directly because the state-space required was rather smaller. Indeed, the strong suppression in the magnitude of the oscillator occupation number as $\Delta$ is increased (there is a reduction by a factor $\sim 100$ in going from $\Delta=0.1$ to $\Delta=0.6$) is the most significant feature in Fig.\ \ref{fig:scaled}a,  which is captured by the $4\Delta^2$ scaling.

Figure \ref{fig:scaled}a also shows that the semi-classical amplitudes provides a very good description of the oscillator occupations for $\Delta\ll 1$. For $\Delta=0.1$ we see that there are small deviations from the semi-classical predictions which become apparent just above threshold and near the bifurcation that occurs at $E_J^{c2}=2.5E_J^c$. As the size of the zero-point fluctuations is increased, these small deviations grow much larger and spread out over a much wider range of $E_J/E_J^c$ values. Nevertheless, the semi-classical amplitude continues to provide a useful estimate of the full quantum results even for $\Delta=0.6$.

We now turn to the fluctuations in the occupation numbers of the oscillators, described by the single mode Fano factors, $F=F_a=F_b$. The value of $F$ decreases progressively the further above threshold we go as shown in Fig.\ \ref{fig:scaled}b. For very small $\Delta$, $F$ is strongly elevated close to threshold (the other side of the peak in $F$ seen below threshold), but decreases rapidly with increasing $E_J/E_J^c$ leading to substantial number state squeezing with $F\sim 0.5$ before the bifurcation at $E_J^{c2}$. For larger $\Delta$ values there is no peak around threshold and $F<1$ throughout though the lowest values are slightly larger than those obtained for very small $\Delta$.

The simple semi-classical analysis in Sec. \ref{sec:scd} can be extended to describe fluctuations (at least to lowest order) by essentially adding a noise term to the equations of motion for the amplitudes, Eqs. \eqref{eq:alpha1} and \eqref{eq:beta1}, so that  they become Langevin equations. Formally, such Langevin equations can be derived within the framework of an approximate semi-classical approach known as the truncated Wigner approximation, as we show in Appendix \ref{sec:twa}. We again make the change to amplitude-phase variables and then linearize about the fixed point values to obtain expressions for the amplitude fluctuations $\langle \delta A^2\rangle=\langle (A-A_0)^2\rangle$ which can be related to the Fano factor in a simple way $F_a\simeq4\langle \delta A^2\rangle$ (details of the calculation are provided in Appendix \ref{sec:twa}).

The comparison of the semi-classical and quantum calculations of the Fano factor shown in Fig.\ \ref{fig:scaled}b shows that the semi-classical Fano factor, which is a function of $E_J/E_J^c$ alone in the symmetric case, can be thought of as giving the low-$\Delta$ limit. As $\Delta$ is increased the deviations from the semi-classical value get stronger around threshold and the bifurcation at $E_J^{c2}=2.5E_J^c$ as well as spreading over a wider range of $E_J/E_J^c$ in much the same way as for the oscillator occupation. Note that the semi-classical calculation predicts a Fano factor which tends to $0.5$ as the system tends to the bifurcation, $E_J\rightarrow E_J^{c2}$. This matches the lowest Fano factors found for the one-oscillator system which occurs as the system tends towards an above-threshold bifurcation at the $2$-photon resonance\,\cite{armour2013}.

\subsection{Asymmetric oscillators}

We now consider what happens when the oscillators are no longer entirely symmetric. We start by considering the case where the zero-point fluctuations of the modes remain the same ($\Delta=\Delta_a=\Delta_b$), but the damping rates are different $r\neq 1$ and then go on to consider the general case where $\Delta_a\neq\Delta_b$ and $r\neq 1$.

\begin{figure}[t]
\centering
{\includegraphics[width=7.0cm]{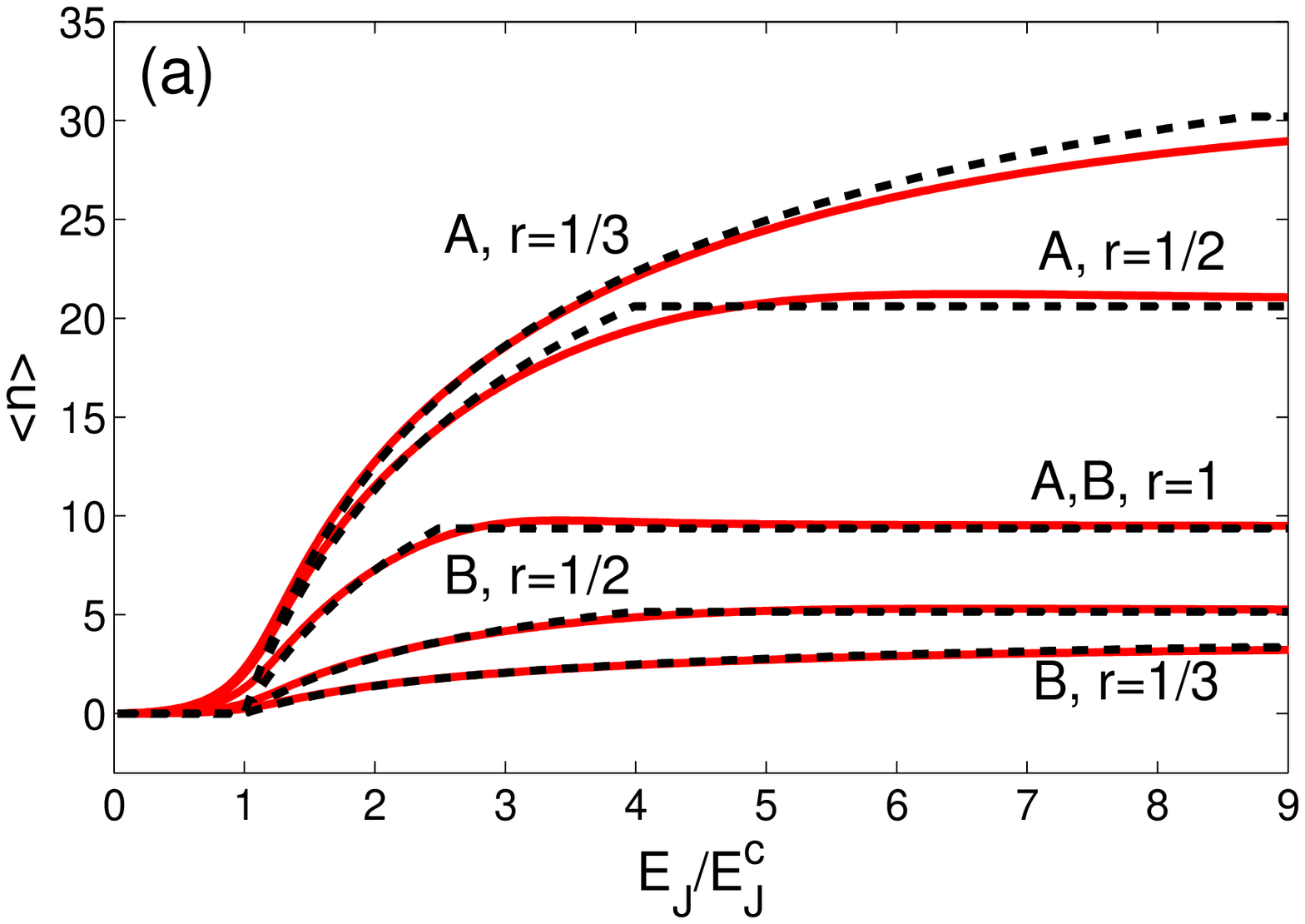}
\includegraphics[width=7.0cm]{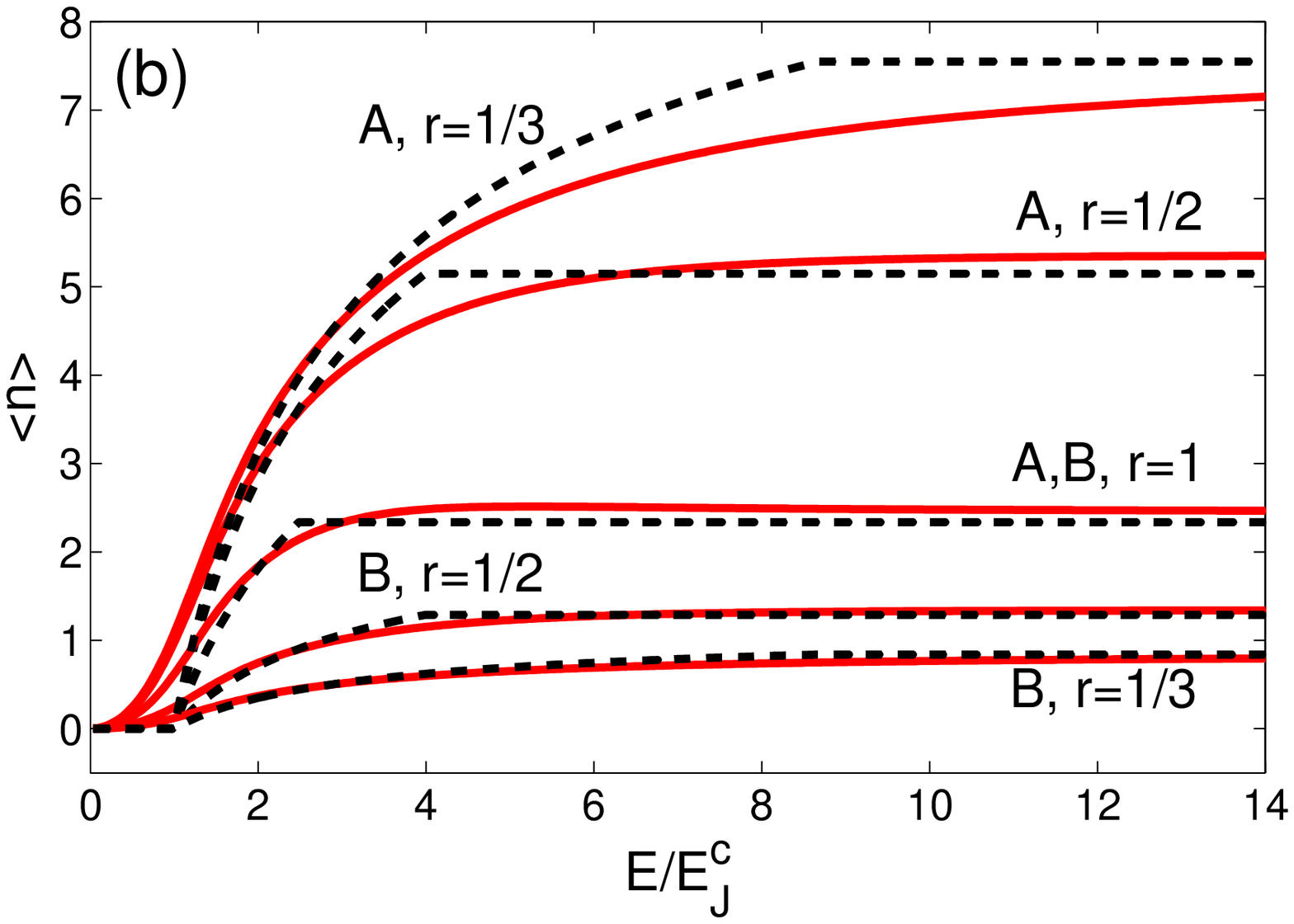}
}
\caption{(Color online) Steady-state occupations $n_a$ and $n_b$ (full lines) compared with the classical values of $A_0^2$ and $B_0^2$ at the stable fixed points (dashed lines) for (a) $\Delta=0.3$ (b) $\Delta=0.6$.  Results are shown for $r=1$, $1/2$ and $r=1/3$ in each case. Note that the semi-classical amplitudes are zero for $E_J<E_J^c$. The above-threshold bifurcation occurs at $E_J^{c2}/E_J^c=2.5$, $4.0$ and $8.7$ for $r=1$, $1/2$ and $r=1/3$, respectively.}
\label{fig:nav}
\end{figure}

\begin{figure}[t]
\centering
{\includegraphics[width=7.0cm]{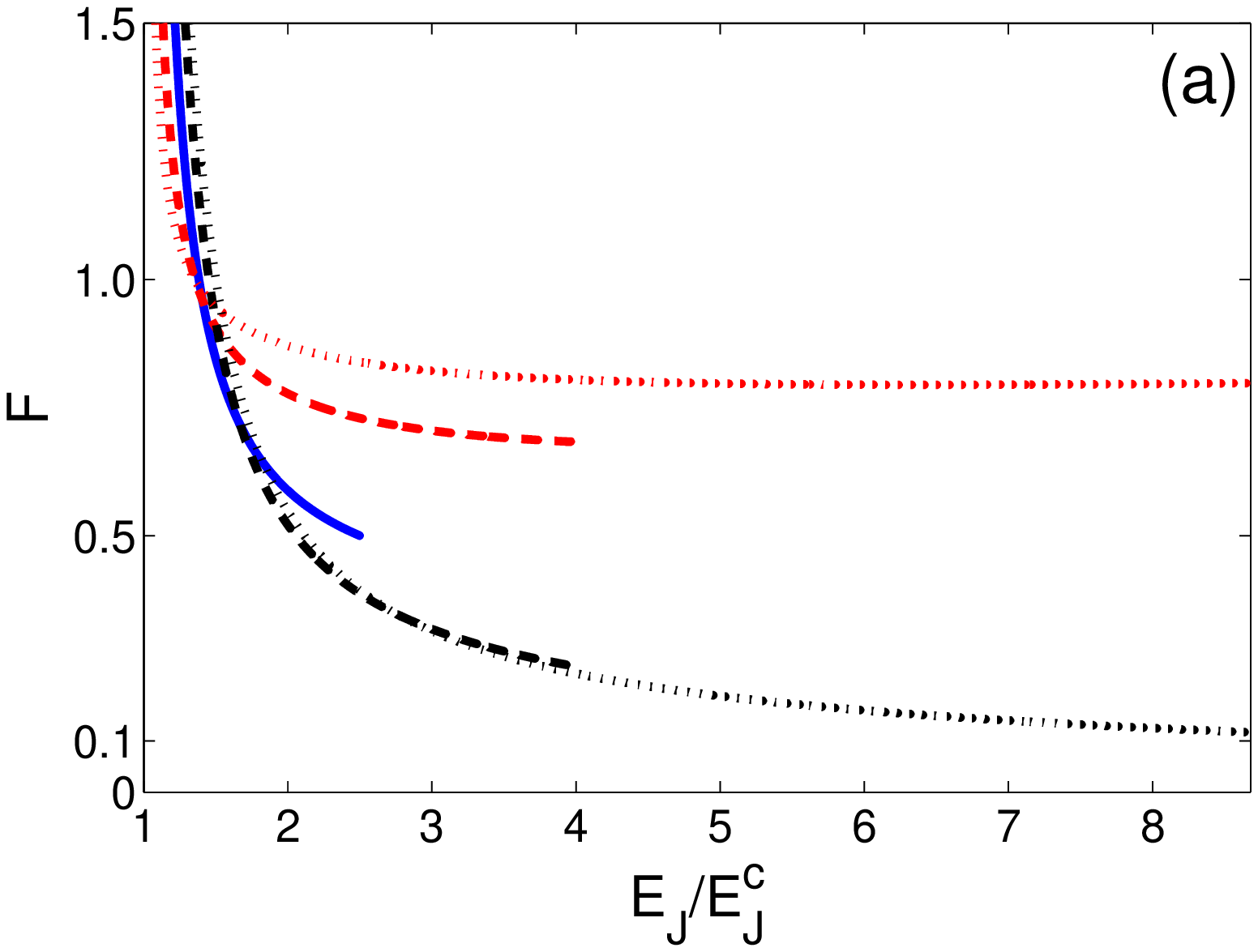}
\includegraphics[width=7.0cm]{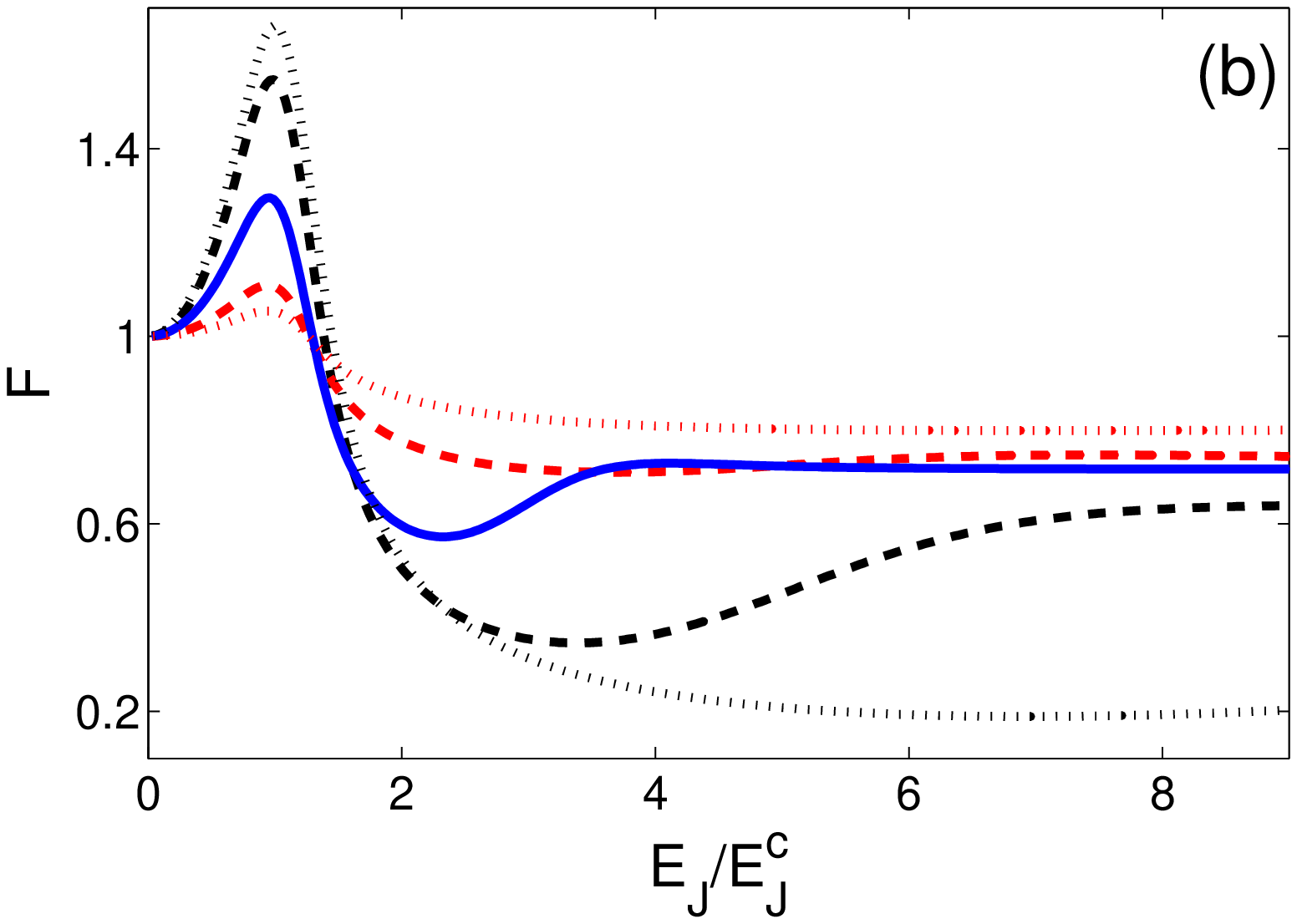}
\includegraphics[width=7.0cm]{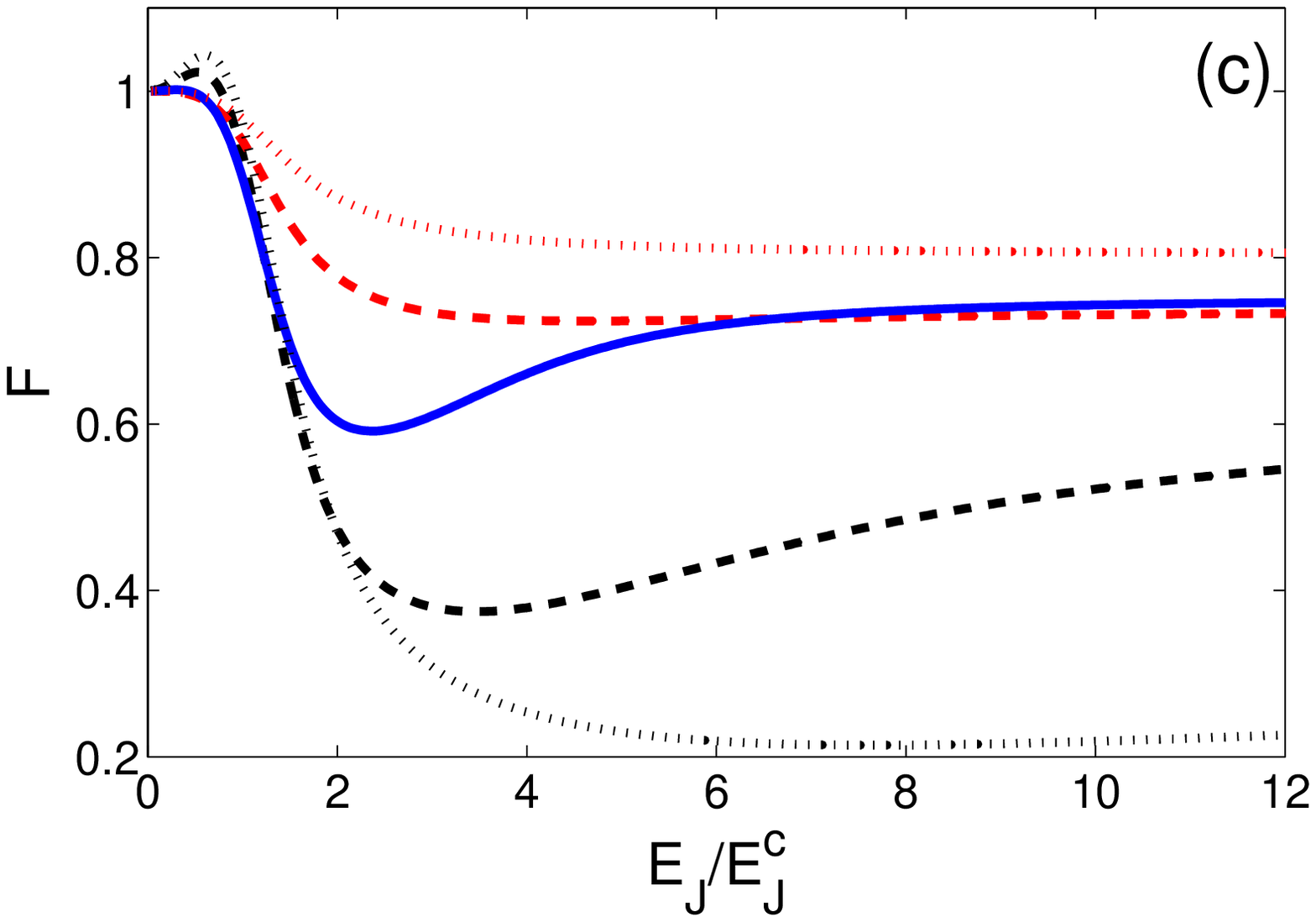}}
\caption{(Color online) Steady-state Fano factors of the modes (a) calculated semi-classically (small-$\Delta$ limit)  and calculated numerically using the master equation for (b) $\Delta=0.3$ and (c) $\Delta=0.6$.  In each case results are shown for $r=1$ (full lines), $r=1/2$ (dashed lines) and $r=1/3$ dotted lines. For $r=1/2$ and $r=1/3$ the upper curves are for oscillator b and the lower ones for oscillator a. The bifurcation occurs at $E_J^{c2}/E_J^c=2.5$, $4.0$ and $8.7$ for $r=1$, $1/2$ and $r=1/3$, respectively. Note that the semi-classical results in (a) are for $E_J^c<E_J<E_J^{c2}$ whilst (b) and (c) cover a broader range of $E_J$ values.}
\label{fig:fano}
\end{figure}

The effect of asymmetric damping on the average occupation numbers of the oscillator (shown in Fig.\ \ref{fig:nav}), is twofold with  both effects following from the underlying semi-classical dynamics discussed in Sec.\ \ref{sec:scd}. Firstly, the bifurcation which occurs at $E_J^{c2}$ is pushed to larger values of $E_J^c$.  Secondly, the average energies of the modes become unequal in proportion to the underlying asymmetry in the damping, $n_b=r^2n_a$.


Figure \ref{fig:fano} shows the effect of asymmetric damping on the occupation number fluctuations for different values of $\Delta$. What is striking here is that the fluctuations become asymmetric and the Fano factor becomes significantly lower than $0.5$ in the less damped oscillator. The lowest values of $F$ are achieved well-above threshold, close to the bifurcation at $E_J^{c2}$ for small-$\Delta$, though for larger $\Delta$ values the minimum $F$ is at a lower value of $E_J$ as the increase in $F$ associated with the bifurcation starts to occur at progressively smaller values of $E_J/E_J^c$ as $\Delta$ is increased. Above the bifurcation the value of $F$ settles down to a steady, but rather higher value.

\begin{figure}[t]
\centering
{\includegraphics[width=7.0cm]{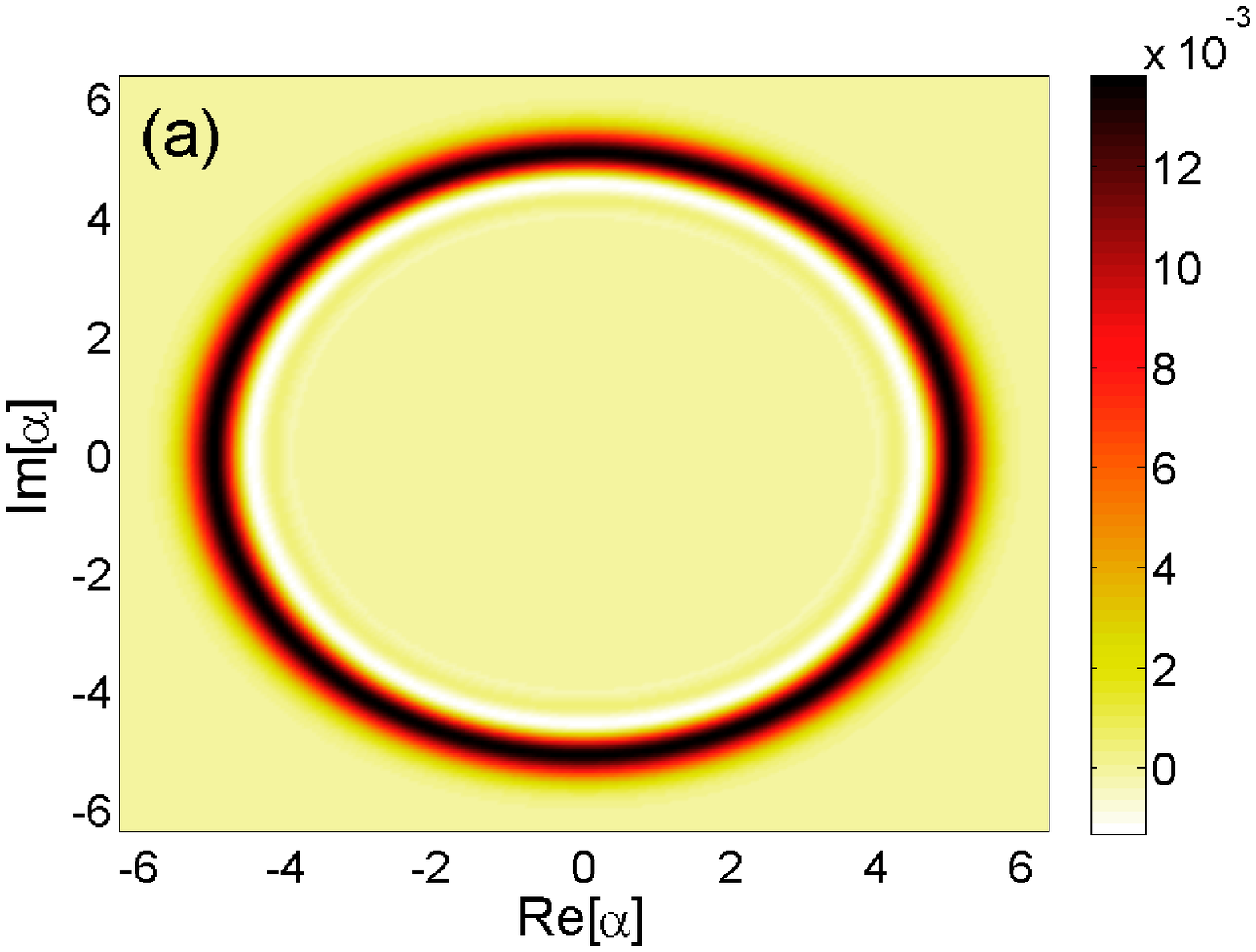}
\includegraphics[width=7.0cm]{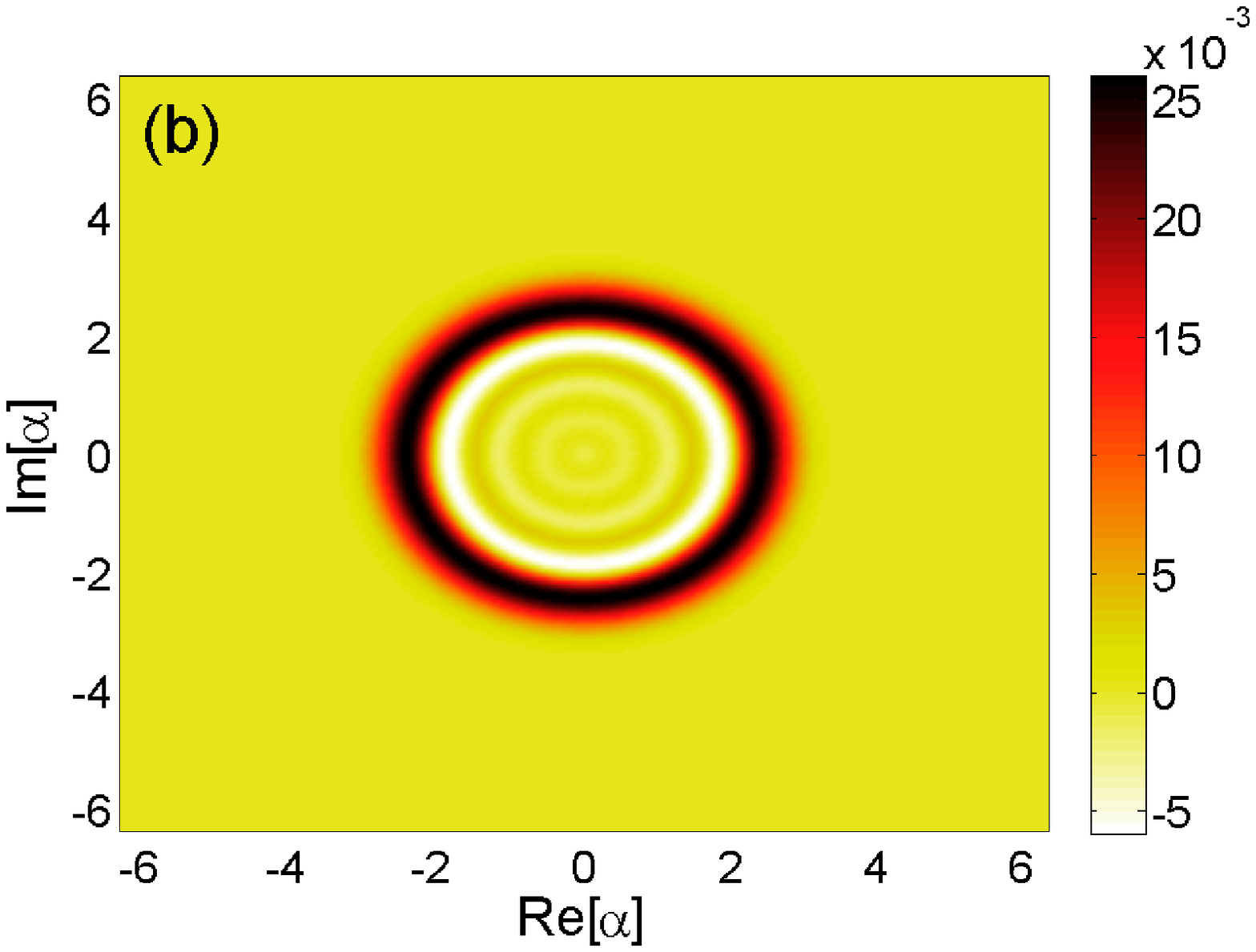}
}
\caption{(Color online) Wigner function of oscillator a for  $r=1/3$, $E_J/E_J^c=6$ and (a) $\Delta=0.3$ (b) $\Delta=0.6$. Negative regions are apparent  in both cases, though more strongly in (b). The Fano factors associated with the states are $F_a=0.19$ (a) and $F_a=0.22$ (b).}
\label{fig:wig}
\end{figure}

The semi-classical calculation predicts a minimum value of $F\simeq 0.1$ for the small-$\Delta$ limit when $r=1/3$, substantially lower than any of the Fano factors predicted for the single-oscillator system\,\cite{armour2013}, and this value continues to decrease for smaller $r$. This  suggests that the asymmetric two-oscillator system may provide a very effective route to preparing a particular mode in a strongly non-classical state at large photon numbers. As $F\rightarrow 0$ the state of the oscillator must eventually become a pure Fock state and so one naturally expects to find negative features in the Wigner function for very small values of $F$. However, the presence of negative regions in a Wigner function is not simply a function of $F$, but also the average occupation number $\langle n\rangle$: as one goes to larger average oscillator occupation numbers, smaller and smaller values of $F$ are required to form negative regions. Figure \ref{fig:wig} illustrates this by showing examples of the Wigner functions for $\Delta=0.3$ and $\Delta=0.6$ with $r=1/3$ and $E_J/E_J^c=6$ where $F\sim 0.2$ in both cases (see Fig.\ \ref{fig:fano}). For $\Delta=0.6$ there is strong evidence of negativity in the Wigner function whilst it is almost washed out for $\Delta=0.3$ since although the Fano factors are very similar, the latter has a much higher average occupation number.

\begin{figure}[t]
\centering
{\includegraphics[width=7.0cm]{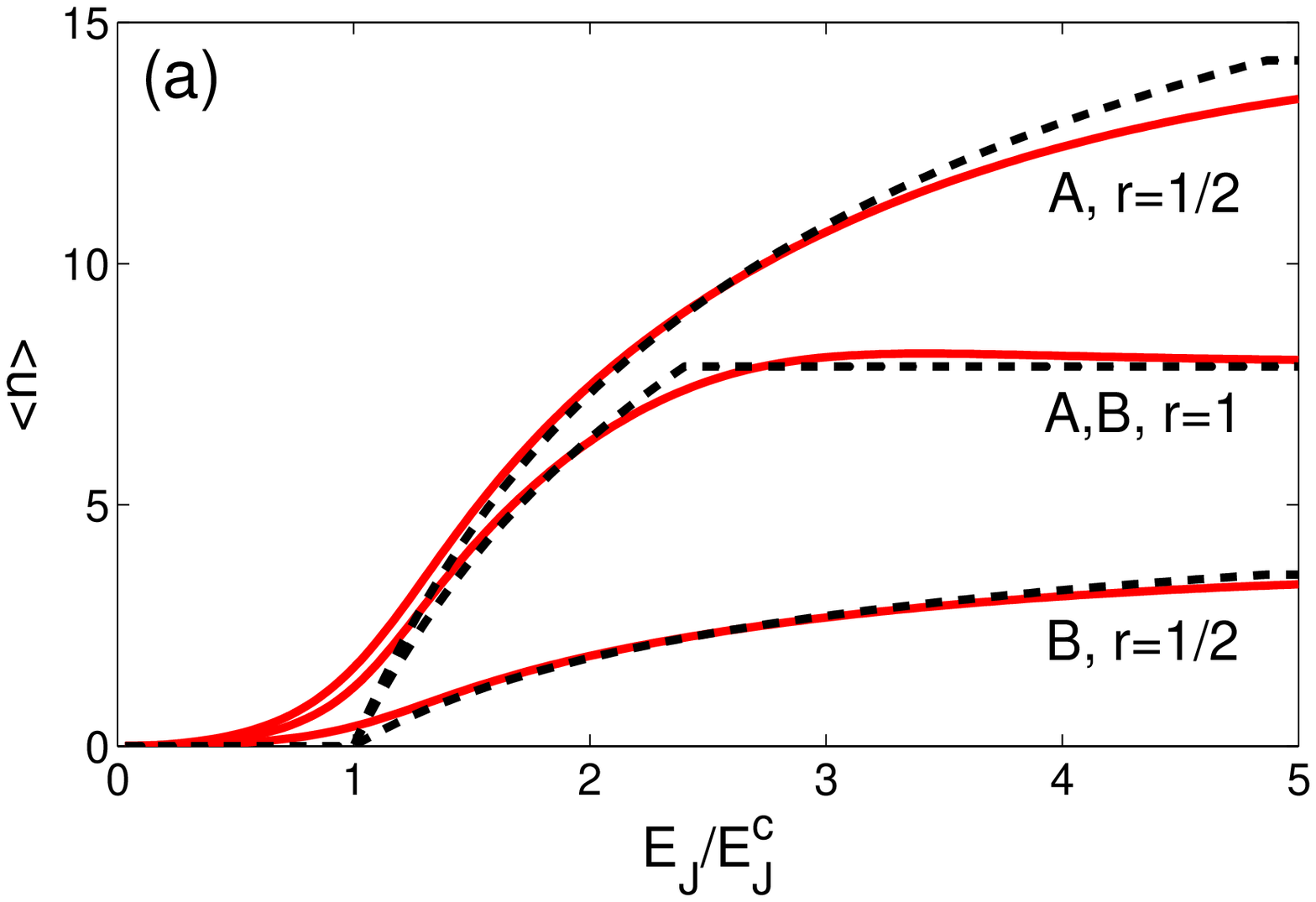}
\includegraphics[width=7.0cm]{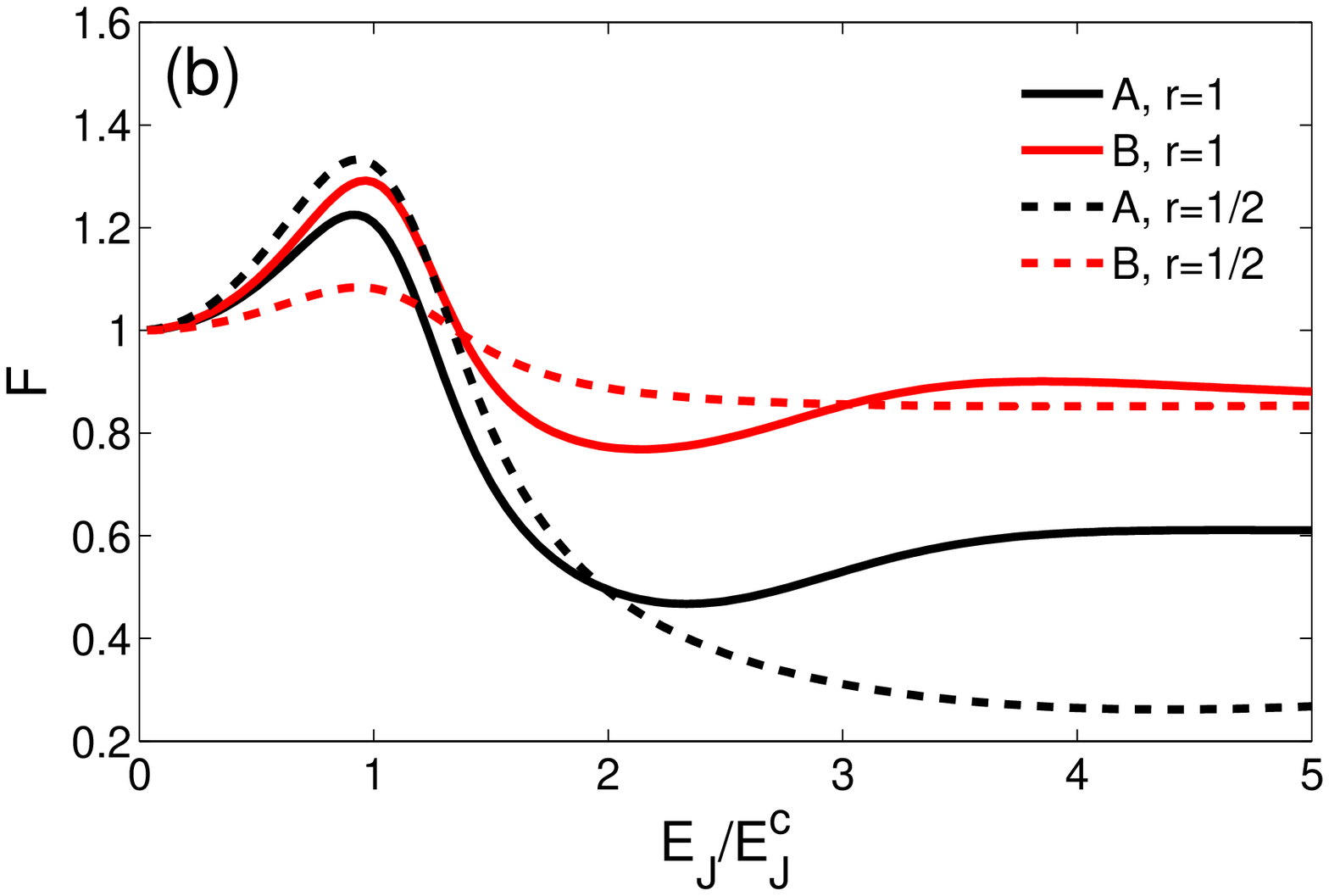}}
\caption{(Color online) Steady-state occupation numbers, $n$, (a) and Fano factors, $F$, (b) of the oscillators for $\Delta_a=0.4$, $\Delta_b=0.2$ with $r=1$ and $1/2$. In $(a)$ the semi-classical predictions are shown as a dotted line and the numerical results as full line in each case.}
\label{fig:fass}
\end{figure}

Finally, we examine the behavior in the regime where $\Delta_a\neq \Delta_b$. Figure \ref{fig:fass} shows examples of the behavior of the occupation numbers and Fano factors of the two oscillators in this case. Interestingly for $r=1$ whilst energy balance means that $n_a=n_b$, the fluctuations in the two modes are no longer the same. When $r\neq 1$ the occupation numbers of the two oscillators spilt according to the usual relation, $n_b=r^2n_a$ and the fluctuations become even more asymmetric. Indeed, the minimum values of the Fano factors, are lower than those in the corresponding cases where $\Delta_a+\Delta_b$ takes the same value, but $\Delta_a=\Delta_b$.

\section{Discussion and Conclusions}

\label{sec:conclude}

We have analyzed the quantum dynamics of two electromagnetic oscillators coupled to a voltage biased Josephson junction. We considered the case where the voltage across the junction was tuned so that the energy lost by a Cooper pair crossing the circuit matches the sum of the photon energies of the two oscillators. In this regime the oscillators are pumped by the flow of Cooper pairs and can become strongly excited. Using a rotating wave approximation, we derived an effective time-independent Hamiltonian for the system and explored the behavior it gives rise to under a wide range of conditions using a mixture of numerical and analytic approaches to solve the master equation. We use a perturbative approach to obtain analytic results for the regime where the occupation of the oscillators is low while in the opposite regime of large occupation numbers a semi-classical approach provides an effective description.

The steady states of the oscillators display signatures of non-classical behavior over a very wide range of conditions with sub-Poissonian photon statistics found in both the low and high occupancy regimes. The strength of the zero-point fluctuations in the oscillators, $\Delta_{a(b)}$, plays an important role: as these are increased the overall excitation level of the oscillators tends to move towards lower photon numbers whilst  the signatures of non-classicality are enhanced. The ratio of the damping rates of the two cavities, described by $r=\sqrt{r_a/r_b}$, also has an interesting effect on the behavior of the system. The photon numbers in the two oscillators are related in a simple way, $n_b=r^2n_a$, as one would expect. However, the quantum fluctuations (e.g.\ measured by the Fano factors $F_{a(b)}$) also become unequal in the asymmetric case, $r\neq 1$. Indeed we find that the Fano factor in the less-damped oscillator can become low enough to lead to significant negative regions in the corresponding Wigner function.

Strong correlations between the two oscillators are to be expected in the regime we consider given the fact that the tunnelling Cooper-pairs excite photons in each of the two oscillators simultaneously. The violation of the classical Cauchy-Schwarz inequality for the photons in the two oscillators, $g^{(2)}_{ab}$, indicates that the corresponding two-mode states are non-classical. It would be natural to also investigate the entanglement between the two oscillators. However, this is complicated by the fact that in practice local voltage fluctuations, even when weak, would be expected to have a very strong influence on phase dependent correlation functions such as $\langle ab\rangle$ which can be important in determining the level of entanglement. This is in contrast to the observables such as photon occupation numbers and correlation functions which we have focussed on here which, as remarked in Sec.\ \ref{sec:qme}, are expected to be only very weakly affected. We plan to address the issue of inter-oscillator entanglement in a future work using a form of the master equation where the effects of voltage fluctuations are explicitly included\,\cite{gramich2013}.

\section*{Acknowledgements}
We thank Miles Blencowe and Fabien Portier for helpful discussions. ADA was supported by the Leverhulme Trust and by EPSRC (UK) under Grant No. EP/I017828. JA and BK acknowledge financial support through the German Science Foundation (DFG) under the SFB/TRR21 and AN336/6-1.

\appendix

\section{Semi-classical calculation of above-threshold fluctuations}
\label{sec:twa}
We can gain useful insights into the dynamics by extending our semi-classical analysis to include quantum fluctuations using a truncated Wigner approximation (TWA)\,\cite{milburn,polkovnikov}. The TWA leads to an approximate equation of motion for the Wigner function of the system, $W(\alpha,\beta)$, in which third-order and higher derivatives are neglected. Dropping higher-order derivatives leads to a Fokker-Planck equation from which we obtain\cite{milburn} Langevin equations for the phase space variables $\alpha$, $\beta$ of the form (for the on-resonance case)
 \begin{eqnarray}
 \dot{\alpha}&=&-\frac{r}{2}\alpha+\frac{i{E}_J}{2\Delta_b{E}_J^c}J_1(2\Delta_b|\beta|)\times \label{eq:alpha2}\\
 &&\left[J_2(2\Delta_a|\alpha|)\frac{\alpha^2\beta}{|\alpha|^2|\beta|}-J_0(2\Delta_a|\alpha|)\frac{\beta^*}{|\beta|}\right] +\eta_{\alpha}(t)
 \nonumber\\
 \dot{\beta}&=&-\frac{1}{2r}\beta+\frac{i{E}_J}{2\Delta_a{E}_J^c}J_1(2\Delta_a|\alpha|)\times  \label{eq:beta2}\\
 &&\left[J_2(2\Delta_b|\beta|)\frac{\beta^2\alpha}{|\beta|^2|\alpha|}-J_0(2\Delta_b|\beta|)\frac{\alpha^*}{|\alpha|}\right] +\eta_{\beta}(t)
\nonumber.
 \end{eqnarray}
The noise terms $\eta_{\alpha(\beta)}(t)$ have zero means and the only non-zero second moments are given by
\begin{eqnarray}
\langle \eta_{\alpha}(t)\eta_{\alpha^*}(t')\rangle&=&\frac{r}{2}\delta(t-t')\\
\langle \eta_{\beta}(t)\eta_{\beta^*}(t')\rangle&=&\frac{1}{2r}\delta(t-t').
\end{eqnarray}
Apart from the noise terms, the equations of motion take the same form\cite{footnotex} as those derived in Sec.\ \ref{sec:scd} [Eqs.\ \eqref{eq:alpha1} and \eqref{eq:beta1}].

We proceed by changing to amplitude and phase variables and then linearizing about the fixed point values, i.e.\ working to first order in $\delta A=A-A_0$, $\delta B=B-B_0$ and $\delta \xi^+=\xi^+-\xi^+_{0}$ with $A_0$, $B_0$, $\xi^+_0$ the fixed point values. For the fixed point just above threshold the amplitude and phase fluctuations become decoupled and on-resonance we find
\begin{equation}
\left (\begin{array}{c}\dot{\delta A}\\ \dot{\delta B}\end{array}\right)=\left (\begin{array}{cc} -\Gamma_a & h_{(a,b)}\\ h_{(b,a)} & -\Gamma_b\end{array}\right)\left (\begin{array}{c}\delta {A}\\  \delta {B}\end{array}\right)+
\left (\begin{array}{c}\eta_{A}\\  \eta_{B}\end{array}\right), \label{eq:vec}
\end{equation}
where
\begin{eqnarray}
\Gamma_{a}&=&\frac{r}{2}+\left(\frac{\Delta_a E_J}{2\Delta_b E_J^c}\right)\times\\
&&J_1(2\Delta_b B_0)\left[J_1(2\Delta_a A_0)+J_3(2\Delta_a A_0)\right]\nonumber\\
\Gamma_{b}&=&\frac{1}{2r}+\left(\frac{\Delta_b E_J}{2\Delta_a E_J^c}\right)\times\\
&&J_1(2\Delta_a A_0)\left[J_1(2\Delta_b B_0)+J_3(2\Delta_b B_0)\right]\nonumber\\
h_{(a,b)}&=&\left(\frac{E_J}{2E_J^c}\right)\left[J_0(2\Delta_b B_0)-J_2(2\Delta_b B_0)\right]\times\\
&&\left[J_0(2\Delta_a A_0)+J_2(2\Delta_a A_0)\right]\nonumber
\end{eqnarray}
and a corresponding expression for $h_{(b,a)}$. The noise terms obey the correlation functions
\begin{eqnarray}
\langle \eta_{A}(t)\eta_{A}(t')\rangle&=&\frac{r}{4}\delta(t-t')\\
\langle \eta_{B}(t)\eta_{B}(t')\rangle&=&\frac{1}{4r}\delta(t-t').
\end{eqnarray}

Using Eq.\ \eqref{eq:vec} we obtain the steady-state variances
\begin{eqnarray}
\langle \delta A^2\rangle&=&\frac{r}{8\Gamma_a}+\frac{h_{(a,b)}}{\Gamma_a}\langle \delta A \delta B\rangle\\
\langle \delta B^2\rangle&=&\frac{1}{8r\Gamma_b}+\frac{h_{(b,a)}}{\Gamma_b}\langle \delta A \delta B\rangle\\
\langle \delta A \delta B\rangle&=& \frac{h_{(a,b)}\Gamma_a/r+h_{(b,a)}\Gamma_br}{8(\Gamma_a+\Gamma_b)(\Gamma_a\Gamma_b-h_{(a,b)}h_{(b,a)})}.
\end{eqnarray}
Recalling that $\alpha$ and $\beta$ are phase space variables of a Wigner function, we can connect these variances to quantum averages:  $\langle A^2\rangle=\langle a^{\dagger}a\rangle+1/2$ and $\langle A^4\rangle=\langle (a^{\dagger}a)^2\rangle+\langle a^{\dagger}a\rangle+1/2$. For fixed points where $A_0\gg 1$, corrections of order $A_0^{-2}$ can be neglected, leading to the simple result,
\begin{eqnarray}
F_{a}&=&\frac{\langle A^4\rangle-\langle A^2\rangle^2-1/4 }{\langle A^2\rangle-1/2}\\
&=&\frac{4A_0^2\langle \delta A^2\rangle+2\langle \delta A^2\rangle^2-1/4}{A_0^2+\langle \delta A^2\rangle-1/2}\\
&\simeq&4\langle \delta A^2\rangle, \label{eq:scfano}
\end{eqnarray}
and there is of course a corresponding relation for $F_b$.

The Langevin equation for  $\delta\xi^+$ takes the form
\begin{equation}
\dot{\delta\xi^+}=-F_+(A_0,B_0)\delta\xi^++\eta_{\xi^+},
\end{equation}
where $\langle \eta_{\xi^+}(t)\eta_{\xi^+}(t')\rangle=2D\delta(t-t')$ with $2D=r/(4A_0^2)+1/(4rB_0^2)$.  Hence we find
\begin{equation}
\langle (\delta\xi^+)^2\rangle=D/F_+(A_0,B_0).
\end{equation}
Note that as the system approaches the bifurcation at $E_J=E_J^{c2}$, $F_+(A_0,B_0)\rightarrow 0$ implying that the total phase fluctuations within this linearized approach diverge.


\begin{thebibliography}{99}
\bibitem{footnotex} In fact the definition of $E_J^c$ is slightly different in Eqs.\ \eqref{eq:alpha2} and \eqref{eq:beta2} as the drift terms in the truncated Wigner approximation are derived from a symmetrically ordered Hamiltonian\,\cite{polkovnikov} (rather than normal order) and hence the factor of ${\rm e}^{(\Delta^2_a+\Delta_b^2)/2}$ does not arise. However, this difference is unimportant in the regime of small $\Delta_a,\Delta_b$ where the TWA is expected to be valid.
\end{thebibliography}
\end{document}